\documentclass[conference]{IEEEtran}
\IEEEoverridecommandlockouts
\usepackage{tikz}
\usepackage{caption}
\usepackage{subcaption}
\usepackage{cite}
\usepackage{amsmath,amssymb,amsfonts}
\usepackage{algorithmic}
\usepackage{graphicx}
\usepackage{textcomp}
\usepackage{xcolor}
\usepackage{color}
\usepackage{colortbl}
\usepackage{array}
\usepackage{pifont}
\usepackage[a4paper, total={184mm,239mm}]{geometry}
\def\BibTeX{{\rm B\kern-.05em{\sc i\kern-.025em b}\kern-.08em
    T\kern-.1667em\lower.7ex\hbox{E}\kern-.125emX}}

\newcommand{\drawsquare}[3]{
    \pgfmathsetmacro\xstart{#1 - #3}
    \pgfmathsetmacro\ystart{#2 - #3}
    \fill[teal] (\xstart,\ystart) rectangle ++(2*#3,2*#3);
}

\newcommand{\drawboundary}{
    \draw[thin] (0,0) rectangle (16,16);
    \foreach \x in {0,1,...,16} {
    }
}

\newcommand{\boxwidth}{.19\linewidth}
\begin{document}

\title{Multi-layer Dataflow: Orchestrate Butterfly Sparsity to Accelerate Attention Computation}
\author{\IEEEauthorblockN{Haibin Wu$^{1}$, Wenming Li$^{1,\ast}$, Kai Yan$^{1,\ast}$, Zhihua Fan$^{1,\ast}$, Peiyang Wu$^{1}$, Yuqun Liu$^{1}$, \\Yanhuan Liu$^{1}$, Ziqing Qiang$^{1}$, Meng Wu$^{1}$, Kunming Zhang$^{1}$, Xiaochun Ye$^{1}$, Dongrui Fan$^{1}$}
\IEEEauthorblockA{$^1$ State Key Lab of Processors, Institute of Computing Technology, CAS, Beijing, China}
}
\maketitle

\vspace{-0.7cm}
\begin{abstract} 
Recent neural networks (NNs) with self attention exhibits competitiveness across different AI domains, but the essential attention mechanism brings massive computation and memory demands. To this end, various sparsity patterns are introduced to reduce the quadratic computation complexity, among which the structured butterfly sparsity has been proven efficient in computation reduction while maintaining model accuracy. However, its addressing inefficiency brings utilization degradation and makes parallelism hard to exploit in general block-oriented architecture like GPU. Since the reconfigurable dataflow accelerators have shown the superiority of better data reusability and architectural flexibility towards general NN-based acceleration, we propose a scalable multilayer dataflow method supported with coarse-grained streaming parallelism, to orchestrate the butterfly sparsity computation on dataflow array. The experiments show that compared with Jetson Xavier NX, our design has a speedup of up to $14.34\times$ ($9.29\times$ on average) as well as $12.3\times$ energy efficiency advancement in attention workloads. In comparison with SOTA butterfly accelerator, our design acquires $1.17\times$ speedup and $3.36\times$ energy efficiency improvement, at the same peak performance.
\end{abstract}

\begin{IEEEkeywords}
attention mechanism, reconfigurable dataflow architecture, structured sparsity, butterfly computation
\end{IEEEkeywords}

\section{Introduction} 
\label{sec:intro}
Prevailing deep learning method, such as \emph{Transformer}, has dominated various domains in Natural Language Processing (NLP) and Computer Vision (CV) \cite{VIT}. Its prominent attention mechanisms\cite{attention} capture comprehensive relationships between tokens and features, through linear mapping implemented by matrix-vector multiplication. However, the computational complexity of attention increases quadratically with the token length, which can extend up to 64K in long sequences, as observed in BERT \cite{BERT}, resulting in substantial computation and memory demands. Various speedup methods with diverse sparsity patterns\cite{Flash,ViTALiTy,Dynamic} have been introduced in the computation process of ${Q,K,V}$ and $softmax(Q(K)^T)V$ to alleviate the computation and memory demands. 

Based on the \emph{theoretical evidence} provided by recent research\cite{monarch,Adaptable}, structured sparsity does not significantly compromise model accuracy, compared with dynamic sparsity\cite{reformer,dota}. Dynamic sparsity poses challenges in general architecture, like GPU due to the inefficiency associated with \emph{randomized addressing}. They often require customized hardware to achieve dynamic element compression or rearrangement within matrices\cite{A3, matRaptor}. In contrast, structured sparsity offers predictable addressing regulations that can be efficiently applied through matrix layout optimization during programming or compiling.

\begin{figure}[tb]
    \centering
    \includegraphics[width=\linewidth,height=0.19\textheight]{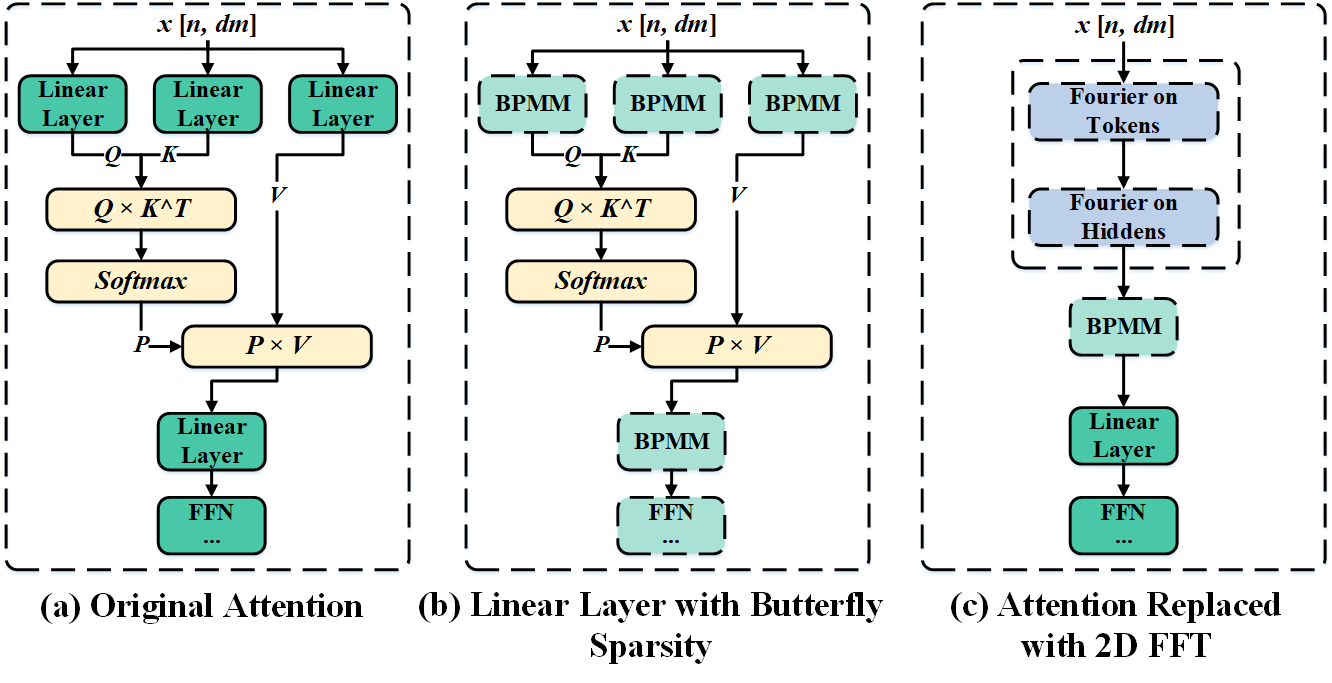}
    \vspace{-0.55cm}
    \captionsetup{font=small}
    \caption{Transformer network with butterfly pattern simplification.}
    \vspace{-0.25cm}
    \label{fig:Transformer_NN}
\end{figure}

Among various structured sparsity, \emph{butterfly sparsity} has been recently proven efficient in enhancing performance of attention workloads while maintaining accuracy across diverse AI tasks\cite{Adaptable}. As shown in Fig.~\ref{fig:Transformer_NN}a\&1b, one of the butterfly sparsity approach is to simplify the linear weight matrices for $Q, K, V$ by using a set of $O(log N)$ butterfly matrices \textbf{$B_{i}$} that have the same sparsity rate of $2/N$ respectively \cite{factor,monarch}. This is a technique we refer to as \textbf{\emph{butterfly pattern matrix-vector multiplication (BPMM)}}. Another approach, such as \emph{FNet}\cite{FNet}, applies \emph{2D Fast Fourier Transform} (2D-FFT) on tokens to heavily decrease the computation amount in attention layers, as illustrated in Fig.~\ref{fig:Transformer_NN}c. With these simplifications, the original dense product is replaced with butterfly computation to reduce the complexity and weight size from $O(N^2)$ to $O(N\log N)$.

\begin{figure}[tb]
    \vspace{-0.05cm}
    \includegraphics[width=\linewidth]{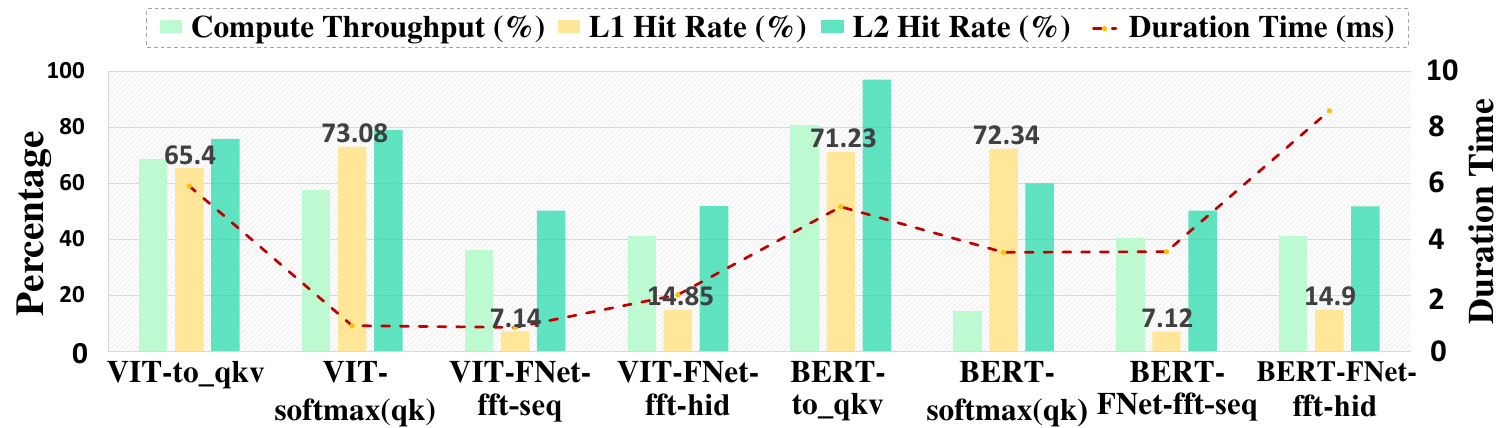}
    \vspace{-0.5cm}
    \captionsetup{font=small}
    \caption{Profiling original dense-based attention and fft-based attention kernels of VIT and BERT on the GPU platform - Jetson Xavier NX.}
    \label{fig:gpu_profiling}
    \vspace{-0.6cm}
\end{figure}

General block-oriented architecture - GPU excels in dense matrix computation, though it lacks energy efficiency, compared to dataflow architectures like TPU\cite{TPU} and Plasticine\cite{plasticine}. In terms of butterfly sparsity, however, the cache-unfriendly addressing results in performance degradation on GPUs. We conducted analyses on NVIDIA - Jetson Xavier NX, by comparing the original dense attention kernels ($to\_qkv$ and $softmax(qk)*v$) and the butterfly attention kernels (\emph{fft-sequence} and \emph{fft-hidden}) of VIT and BERT. Detailed profiling results are presented in Fig.~\ref{fig:gpu_profiling}. Although the data throughput is given in \emph{adequate batch parallelism} of 128 in GPU, the \emph{hit rates}, especially for the L1 cache, exhibit significant degradation in the FFT-based kernels compared to the dense $q,k,v$ kernels in both VIT and BERT. Furthermore, their overall performance indicated by duration time fails to exhibit obvious speedup (even a reduced performance for BERT in large scales), despite the \emph{theoretical reduction of computational complexity} mentioned above with the butterfly sparsity.

 Aiming for tackling the challenges of \textbf{accessing inefficiency} and \textbf{parallelism exploitation hardness} in structured butterfly sparsity computation, we introduce a \emph{scalable multilayer dataflow orchestration} method applied on a flexible reconfigurable dataflow architecture, to speed up butterfly computation for attention workloads. Our contributions are as follows:

\begin{itemize}
    \item \textbf{Algorithm aspects:}\item \textbf{\underline{i}.} We propose a \emph{multilayer dataflow method} to tackle the hardness of orchestrate the data reuse of butterfly sparsity on a \emph{reconfigurable dataflow architecture substrate}. \textbf{\underline{ii}.} we equip this dataflow substrate with \emph{block-level DFG node scheduling} as well as the decoupled function unit design, to exploit effective \emph{coarse-grained streaming parallelism}. 
    
    \item \textbf{\underline{iii}.} We introduce a \emph{multi-stage division method} based on the \emph{Cooley-Tukey algorithm} to enhance the scalability of our dataflow approach towards butterfly sparsity in various data scales. \textbf{\underline{iv}.} Moreover, we architect the scratchpad memory with a multi-line design to implement \emph{transpose-free SIMD parallelism} in large-scale vector sequence for boosting the overall computation performance.
    
    \item Experiment results demonstrate the superiority of the proposed multilayer dataflow method with outstanding speedup and energy efficiency in attention workloads over the advanced GPU and state-of-the-art accelerators.
\end{itemize}

\section{Background and Related Work}
\label{sec:bg+motivation}
In this session, we begin by discussing the existing variants of sparsity in attention workloads and emphasize the superiority of butterfly sparsity for global relationships capturing. Next, we summarize the shortcomings in recent architectural works towards butterfly sparsity computation.

\begin{figure}[tb]
    \centering
    \includegraphics[width=\linewidth]{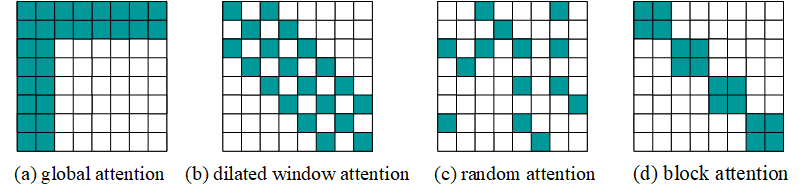}
    \vspace{-0.5cm}
    \captionsetup{font=small}
    \caption{Fundamental patterns in existing sparsity variants.}
    \label{fig:basic_sparsity}
    \vspace{-0.4cm}
\end{figure}

\subsection{Sparsity Variants for Attention Computation} 
The prominent self-attention mechanism effectively facilitates interaction among tokens and features, overcoming the limitations of CNNs and RNNs. But it is plagued by quadratic computation and memory requirements. Recent sparse attention approaches introduce diverse sparsity patterns in attention matrices or linear weighted matrices, as shown in Fig.~\ref{fig:basic_sparsity}. One notable implementation is dynamic sparsity, which is determined by the \emph{sequence content}. Kitaev \emph{et al.} \cite{reformer} proposed Locality Sensitive Hashing (LSH) to infer content-based sparsity. Qu \emph{et al}. utilized low-rank linear transformations to estimate attention scores \cite{dota} and select the sparse pattern dynamically through top-k selection. However, these content-based methods scale poorly in long sequence and may be \emph{application-fixed and awkward} when facing with variable sparse density.

Static structured sparsity as an alternative method addresses these drawbacks by applying regular sparsity patterns to attention and linear matrices. Common structured approaches often \emph{combine several basic patterns} shown in Fig.~\ref{fig:basic_sparsity}.(a,b,d). For instance, Child \emph{et al}. \cite{generating} introduced a sparsity model combining sliding and dilated sliding window pattern, and acquired state-of-the-art accuracy results in long sequences. Also, in Longformer \cite{longformer}, Beltagy \emph{et al}. proposed the global-plus-sliding-window pattern that achieved linear scalability of sequence length while maintaining high accuracy in long document tasks. These combination methods aim to improve patterns capturing within attention matrix, as using only one partial sparsity has \emph{limited expression capability} for \emph{token relationships} and \emph{weight features}.

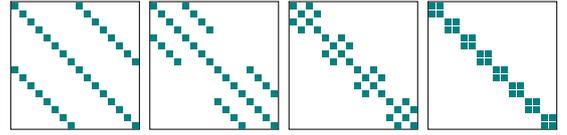
\begin{figure}[tb]
    \centering
\resizebox{\boxwidth}{!}{
\begin{tikzpicture}
    \drawboundary
    
    \drawsquare{ 0.5}{ 7.5}{0.4}
    \drawsquare{ 1.5}{ 6.5}{0.4} 
    \drawsquare{ 2.5}{ 5.5}{0.4} 
    \drawsquare{ 3.5}{ 4.5}{0.4} 
    \drawsquare{ 4.5}{ 3.5}{0.4} 
    \drawsquare{ 5.5}{ 2.5}{0.4} 
    \drawsquare{ 6.5}{ 1.5}{0.4} 
    \drawsquare{ 7.5}{ 0.5}{0.4} 

    \drawsquare{ 0.5}{15.5}{0.4}
    \drawsquare{ 1.5}{14.5}{0.4}
    \drawsquare{ 2.5}{13.5}{0.4}
    \drawsquare{ 3.5}{12.5}{0.4}
    \drawsquare{ 4.5}{11.5}{0.4}
    \drawsquare{ 5.5}{10.5}{0.4}
    \drawsquare{ 6.5}{ 9.5}{0.4}
    \drawsquare{ 7.5}{ 8.5}{0.4}
    \drawsquare{ 8.5}{ 7.5}{0.4}
    \drawsquare{ 9.5}{ 6.5}{0.4}
    \drawsquare{10.5}{ 5.5}{0.4}
    \drawsquare{11.5}{ 4.5}{0.4}
    \drawsquare{12.5}{ 3.5}{0.4}
    \drawsquare{13.5}{ 2.5}{0.4}
    \drawsquare{14.5}{ 1.5}{0.4}
    \drawsquare{15.5}{ 0.5}{0.4}

    \drawsquare{ 8.5}{15.5}{0.4} 
    \drawsquare{ 9.5}{14.5}{0.4} 
    \drawsquare{10.5}{13.5}{0.4} 
    \drawsquare{11.5}{12.5}{0.4} 
    \drawsquare{12.5}{11.5}{0.4} 
    \drawsquare{13.5}{10.5}{0.4} 
    \drawsquare{14.5}{ 9.5}{0.4} 
    \drawsquare{15.5}{ 8.5}{0.4}    
\end{tikzpicture}}
\resizebox{\boxwidth}{!}{
\begin{tikzpicture}
	\drawboundary
    
    \drawsquare{ 0.5}{11.5}{0.4}
    \drawsquare{ 1.5}{10.5}{0.4} 
    \drawsquare{ 2.5}{ 9.5}{0.4} 
    \drawsquare{ 3.5}{ 8.5}{0.4} 
    
    \drawsquare{ 8.5}{ 3.5}{0.4} 
    \drawsquare{ 9.5}{ 2.5}{0.4} 
    \drawsquare{10.5}{ 1.5}{0.4} 
    \drawsquare{11.5}{ 0.5}{0.4} 

    \drawsquare{ 0.5}{15.5}{0.4}
    \drawsquare{ 1.5}{14.5}{0.4}
    \drawsquare{ 2.5}{13.5}{0.4}
    \drawsquare{ 3.5}{12.5}{0.4}
    \drawsquare{ 4.5}{11.5}{0.4}
    \drawsquare{ 5.5}{10.5}{0.4}
    \drawsquare{ 6.5}{ 9.5}{0.4}
    \drawsquare{ 7.5}{ 8.5}{0.4}
    \drawsquare{ 8.5}{ 7.5}{0.4}
    \drawsquare{ 9.5}{ 6.5}{0.4}
    \drawsquare{10.5}{ 5.5}{0.4}
    \drawsquare{11.5}{ 4.5}{0.4}
    \drawsquare{12.5}{ 3.5}{0.4}
    \drawsquare{13.5}{ 2.5}{0.4}
    \drawsquare{14.5}{ 1.5}{0.4}
    \drawsquare{15.5}{ 0.5}{0.4}

    \drawsquare{ 4.5}{15.5}{0.4}
    \drawsquare{ 5.5}{14.5}{0.4}
    \drawsquare{ 6.5}{13.5}{0.4}
    \drawsquare{ 7.5}{12.5}{0.4}
    
    \drawsquare{12.5}{ 7.5}{0.4}
    \drawsquare{13.5}{ 6.5}{0.4}
    \drawsquare{14.5}{ 5.5}{0.4}
    \drawsquare{15.5}{ 4.5}{0.4}   
\end{tikzpicture}}
\resizebox{\boxwidth}{!}{
\begin{tikzpicture}
	\drawboundary
    
    \drawsquare{ 0.5}{13.5}{0.4}
    \drawsquare{ 1.5}{12.5}{0.4} 
    \drawsquare{ 4.5}{ 9.5}{0.4} 
    \drawsquare{ 5.5}{ 8.5}{0.4} 
    \drawsquare{ 8.5}{ 5.5}{0.4} 
    \drawsquare{ 9.5}{ 4.5}{0.4} 
    \drawsquare{12.5}{ 1.5}{0.4} 
    \drawsquare{13.5}{ 0.5}{0.4} 

    \drawsquare{ 0.5}{15.5}{0.4}
    \drawsquare{ 1.5}{14.5}{0.4}
    \drawsquare{ 2.5}{13.5}{0.4}
    \drawsquare{ 3.5}{12.5}{0.4}
    \drawsquare{ 4.5}{11.5}{0.4}
    \drawsquare{ 5.5}{10.5}{0.4}
    \drawsquare{ 6.5}{ 9.5}{0.4}
    \drawsquare{ 7.5}{ 8.5}{0.4}
    \drawsquare{ 8.5}{ 7.5}{0.4}
    \drawsquare{ 9.5}{ 6.5}{0.4}
    \drawsquare{10.5}{ 5.5}{0.4}
    \drawsquare{11.5}{ 4.5}{0.4}
    \drawsquare{12.5}{ 3.5}{0.4}
    \drawsquare{13.5}{ 2.5}{0.4}
    \drawsquare{14.5}{ 1.5}{0.4}
    \drawsquare{15.5}{ 0.5}{0.4}

    \drawsquare{ 2.5}{15.5}{0.4} 
    \drawsquare{ 3.5}{14.5}{0.4} 
    \drawsquare{ 6.5}{11.5}{0.4} 
    \drawsquare{ 7.5}{10.5}{0.4} 
    \drawsquare{10.5}{ 7.5}{0.4} 
    \drawsquare{11.5}{ 6.5}{0.4} 
    \drawsquare{14.5}{ 3.5}{0.4} 
    \drawsquare{15.5}{ 2.5}{0.4}    
\end{tikzpicture}}
\resizebox{\boxwidth}{!}{
\begin{tikzpicture}
	\drawboundary

    \drawsquare{ 0.5}{15.5}{0.4}
    \drawsquare{ 0.5}{14.5}{0.4} 
    \drawsquare{ 1.5}{15.5}{0.4} 
    \drawsquare{ 1.5}{14.5}{0.4} 
    \drawsquare{ 2.5}{13.5}{0.4}
    \drawsquare{ 2.5}{12.5}{0.4} 
    \drawsquare{ 3.5}{13.5}{0.4} 
    \drawsquare{ 3.5}{12.5}{0.4} 
    \drawsquare{ 4.5}{11.5}{0.4}
    \drawsquare{ 4.5}{10.5}{0.4}
    \drawsquare{ 5.5}{11.5}{0.4} 
    \drawsquare{ 5.5}{10.5}{0.4} 
    \drawsquare{ 6.5}{ 9.5}{0.4} 
    \drawsquare{ 6.5}{ 8.5}{0.4} 
    \drawsquare{ 7.5}{ 9.5}{0.4} 
    \drawsquare{ 7.5}{ 8.5}{0.4} 
    \drawsquare{ 8.5}{ 7.5}{0.4} 
    \drawsquare{ 8.5}{ 6.5}{0.4}
    \drawsquare{ 9.5}{ 7.5}{0.4} 
    \drawsquare{ 9.5}{ 6.5}{0.4} 
    \drawsquare{10.5}{ 5.5}{0.4} 
    \drawsquare{10.5}{ 4.5}{0.4} 
    \drawsquare{11.5}{ 5.5}{0.4} 
    \drawsquare{11.5}{ 4.5}{0.4} 
    \drawsquare{12.5}{ 3.5}{0.4} 
    \drawsquare{12.5}{ 2.5}{0.4} 
    \drawsquare{13.5}{ 3.5}{0.4} 
    \drawsquare{13.5}{ 2.5}{0.4} 
    \drawsquare{14.5}{ 1.5}{0.4} 
    \drawsquare{14.5}{ 0.5}{0.4} 
    \drawsquare{15.5}{ 1.5}{0.4} 
    \drawsquare{15.5}{ 0.5}{0.4} 
\end{tikzpicture}}
    \captionsetup{font=small}
    \vspace{+0.1cm}
    \caption{The consecutive weight matrices multiplication of butterfly sparsity in diverse stride patterns.}
    \label{fig:butterfly_factor_matrices}
    \vspace{-0.4cm}
\end{figure}

\begin{figure*}[t]
    \vspace{-0.2cm}
    \centering
    \hspace*{\fill}
    \begin{subfigure}{0.67\linewidth}
    \includegraphics[width=\linewidth]{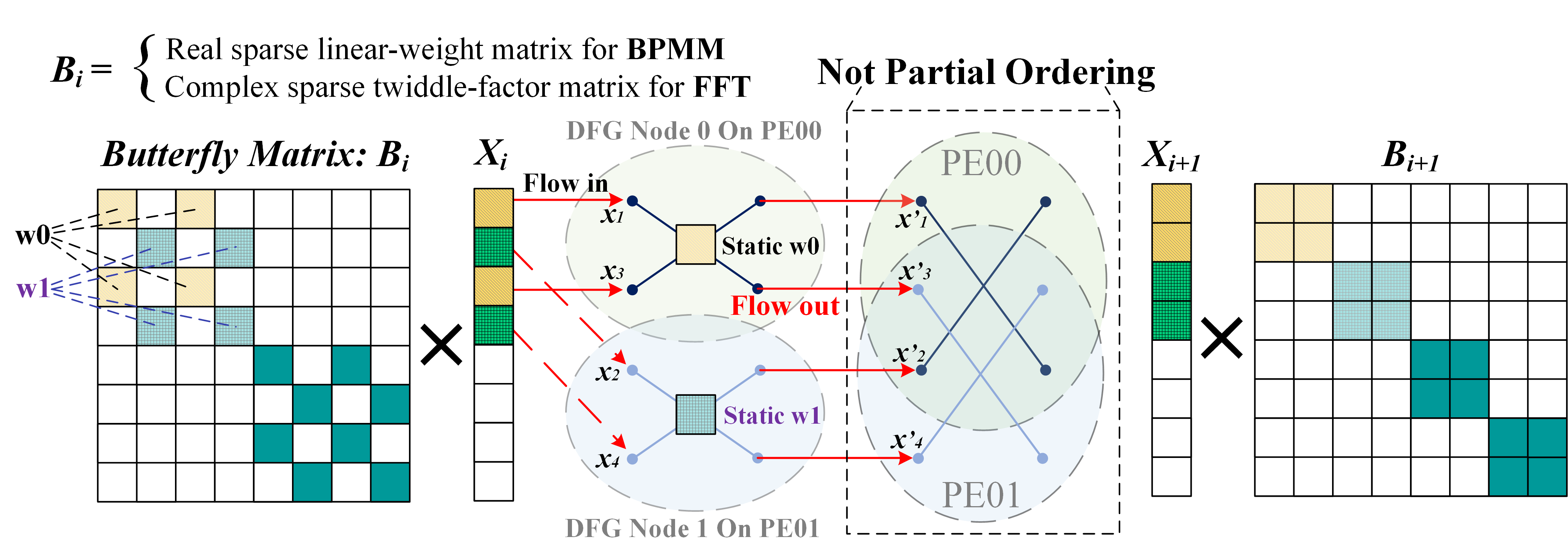}
    \caption{Original butterfly dataflow between DFG nodes is not in partial-order relation because of the mutual data dependence for element swapping.}
    \label{fig:butterfly_before}
    \end{subfigure}    
    \hfill
    \begin{subfigure}{0.25\linewidth}
    \includegraphics[width=\linewidth]{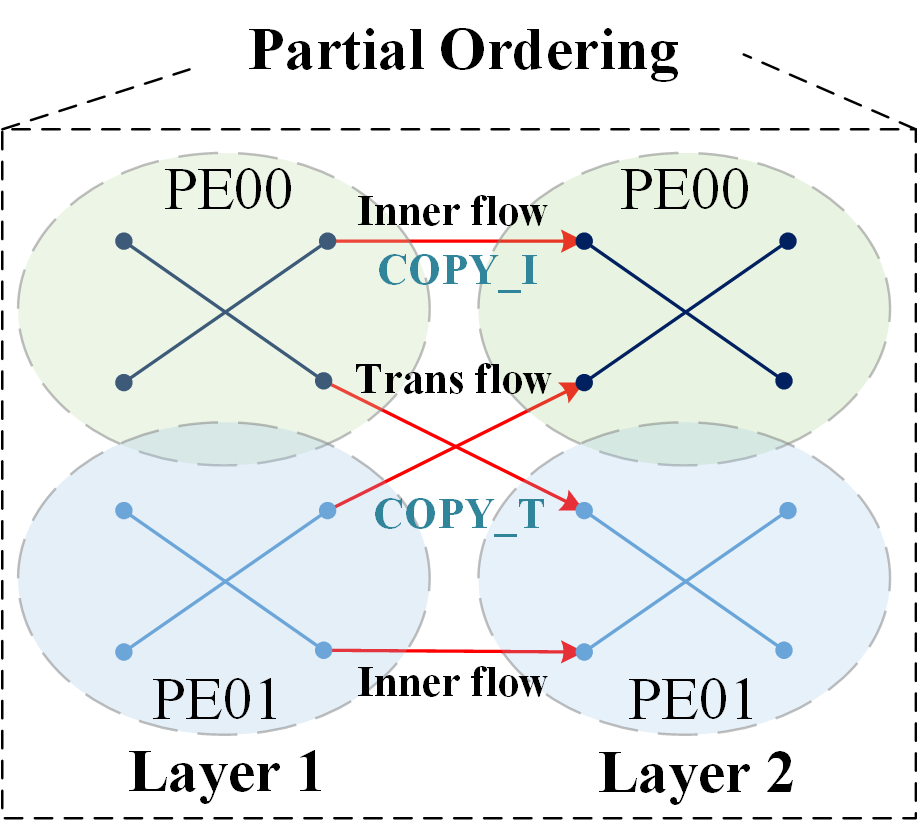}
    \caption{Partial-order dataflow dependence reconstructed in layers.}
    \label{fig:butterfly_after}
    \end{subfigure}
    \hspace*{\fill}
    \caption{Butterfly sparsity computation expressed as the dataflow execution with data dependence in multiple layers.}
    \label{fig:butterfly_swap}
    \vspace{-0.5cm}
\end{figure*}

\subsection{Principle of Butterfly Matrices and Fourier Transform} 

Butterfly sparsity is a special global pattern \textbf{beyond the restriction of limited relationship}. In Fig.~\ref{fig:butterfly_factor_matrices}, it involves a set of regular structured matrices whose product - \emph{BPMM} captures the overall relationships of tokens or features in a binary reductive way with logarithmic complexity. \textbf{\underline{A}.} The first sparse approach is applying these matrices to the linear layers of ${Q,K,V}$ and \textbf{\emph{feed forward network}} (FFN), correspond to Fig.~\ref{fig:Transformer_NN}b. In the works\cite{factor,monarch}, decomposition methods that represent linear weights as the multiplication of butterfly matrices are proposed and \emph{proven effective in parameter simplification}. \textbf{\underline{B}.} The second further sparse approach in attention workloads is applying \emph{Fourier Transform} to replace or approximate computation of \textbf{\emph{attention matrix}}\cite{FNet, Fourier_Transformer}, correspond to Fig.~\ref{fig:Transformer_NN}c. The principle of \emph{Discrete Fourier Transform} (DFT) is illustrated in Eq~\eqref{eq:dft}. The result vector $X$ is obtained by performing matrix multiplication between an input vector $x$ and a dense matrix $\Omega$ whose entry at row $k$ and column $n$ is represented as $\omega_N^{kn}$.

\vspace{-0.1cm}
\begin{equation} \label{eq:dft}
  \resizebox{0.7\linewidth}{!}{$
  X_{k} = \sum_{n=0}^{N-1} x_n \omega_{N}^{kn}, \ k = 0, \dots, N-1.
  $}
\end{equation}
\vspace{-0.25cm}

A famous fast algorithm for DFT computing is the \emph{\textbf{Cooley-Tukey decimation-in-time}} FFT \cite{Cooley1965}, which recursively divides the sequence into odd and even subsequences to reduce the computation from $O(N^2)$ to $O(N\log N)$, as shown in Eq~\eqref{eq:fft}.

\vspace{-0.2cm}
\begin{equation} \label{eq:fft}
  \resizebox{0.7\linewidth}{!}{$
\begin{aligned}
  X_{k} = \sum_{n=0}^{\frac{N}{2}-1} x_{2n} \omega_{\frac{N}{2}}^{kn} + \omega_N^k\sum_{n=0}^{\frac{N}{2}-1} x_{2n+1} \omega_{\frac{N}{2}}^{kn}, \\
  X_{k+\frac{N}{2}} = \sum_{n=0}^{\frac{N}{2}-1} x_{2n} \omega_{\frac{N}{2}}^{kn} - \omega_N^k\sum_{n=0}^{\frac{N}{2}-1} x_{2n+1} \omega_{\frac{N}{2}}^{kn}.
\end{aligned}
$}
\end{equation}
\vspace{-0.1cm}

It can be expressed in a matrix form:
\begin{equation} \label{eq:fft_matrix}
  \resizebox{0.9\linewidth}{!}{$
\begin{aligned}
  X = \Omega_N x &= 
  \begin{bmatrix} 
  \Omega_{N/2} x_{\mathrm{even}} + D_{N/2} \Omega_{N/2} x_{\mathrm{odd}} \\ \Omega_{N/2} x_{\mathrm{even}} - D_{N/2} \Omega_{N/2} x_{\mathrm{odd}} \end{bmatrix} \\
  &= \begin{bmatrix}
  I_{N/2} & D_{N/2} \\
  I_{N/2} & -D_{N/2}
  \end{bmatrix}
  \begin{bmatrix}
  \Omega_{N/2} & 0 \\
             0 & \Omega_{N/2}
  \end{bmatrix}
  \begin{bmatrix}
  x_{\mathrm{even}} \\
  x_{\mathrm{odd}}
  \end{bmatrix} \\
  &= B_N
  \begin{bmatrix}
  \Omega_{N/2} & 0 \\
             0 & \Omega_{N/2}
  \end{bmatrix}
  P_N x,
\end{aligned}
$}
\end{equation}
\vspace{-0.1cm}where $D_{N/2}$ is a diagonal matrix with entries $1, \omega_N^1, \omega_N^2, \dots, \omega_N^{N/2 - 1}$. $B_N$ is a $2\times2$ block matrix consisting of diagonal matrices referred to as \emph{“butterfly factors”}~\cite{monarch}. $P_N$ represents a permutation matrix responsible for performing a sequence reversion.

The middle matrix $\Omega$ can be defined recursively, with the base case being $\Omega_{N/N} = [\omega_1^{00}] = [1]$. 
Expanding the recursion as follows:
\begin{equation} \label{eq:fft-factorization}
\resizebox{0.9\linewidth}{!}{$
\begin{aligned}
  \Omega_N &= B_N \begin{bmatrix} \Omega_{N/2} & 0 \\ 0 & \Omega_{N/2} \end{bmatrix} P_N \\
      &= B_N \begin{bmatrix} B_{N/2} & 0 \\ 0 & B_{N/2} \end{bmatrix}
         \begin{bmatrix} \Omega_{N/4} & 0 & 0 & 0 \\ 0 & \Omega_{N/4} & 0 & 0 \\ 0 & 0 & \Omega_{N/4} & 0 \\ 0 & 0 & 0 & \Omega_{N/4} \end{bmatrix}
         \begin{bmatrix} P_{N/2} & 0 \\ 0 & P_{N/2} \end{bmatrix} P_N \\
      &= \cdots \\
      &= \left( B_N \dots \begin{bmatrix} B_{2} & \dots & 0 \\ \vdots & \ddots & \vdots \\ 0 & \dots & B_2 \end{bmatrix} \right)
         \left( \begin{bmatrix} P_{2} & \dots & 0 \\ \vdots & \ddots & \vdots \\ 0 & \dots & P_2 \end{bmatrix} \dots P_N \right). 
\end{aligned}
$}
\end{equation}

The consecutive multiplying of $log N$ sparse matrices $B_i$ on the left in Eq~\eqref{eq:fft-factorization} shares the same regular pattern as BPMM in Fig.~\ref{fig:butterfly_factor_matrices}. But it is calculated in complex values which result in different \emph{arithmetic density} (ratio between calculation and accessing) of vector elements, compared with the real-value BPMM.
\vspace{-0.1cm}
\subsection{Butterfly Computation Acceleration} 
The parallelism of butterfly sparsity is not easy to exploit on block-oriented architectures like GPU, because of the incremental stride patterns in the consecutive matrices $B_i$ as shown in Fig.~\ref{fig:butterfly_factor_matrices}. Fan \emph{et al}. proposed TensorFHE\cite{TensorFHE} by using \emph{Tensor Cores} to boost the overall performance for butterfly sparsity computing, but it didn't make any compression on sparse elements and actually introduced a mass of \emph{redundant computation} with poor energy efficiency. MaPU\cite{MaPU} introduced multi-granularity parallel memory banks to implement flexible vector splicing and the diversity of accessing strides, but the data shuffling latency remains unsolved between different butterfly stages. The work\cite{Adaptable} proposed butterfly sparsity network with \emph{comprehensive accuracy evaluations} as well as a software and hardware co-design to speed up the attention workloads, but it is fixed as a customized architecture with single concatenation on butterfly acceleration, lacking of \textbf{\emph{architectural reconfigurability}}. Other customized designs using pipeline-stage parallelism \cite{fft_survey} also \emph{lack the flexibility} for other various tensor-based operations in neural networks (NNs) besides the butterfly computation.
\section{Dataflow Architecture Orchestration for Butterfly Sparsity} 
\label{sec:design}
In this section, we first introduce the substrate of our reconfigurable dataflow architecture as the basis of our orchestration method. Based on the accessing inefficiency in butterfly matrices computation as profiled in Fig.~\ref{fig:gpu_profiling}, we introduce layer-level flowing operation in DFG for butterfly element swapping and construct a multilayer DFG to better orchestrate the data reuse in consecutive butterfly matrices multiplications.

\begin{figure}[tp]
    \centering
    \includegraphics[width=\linewidth]{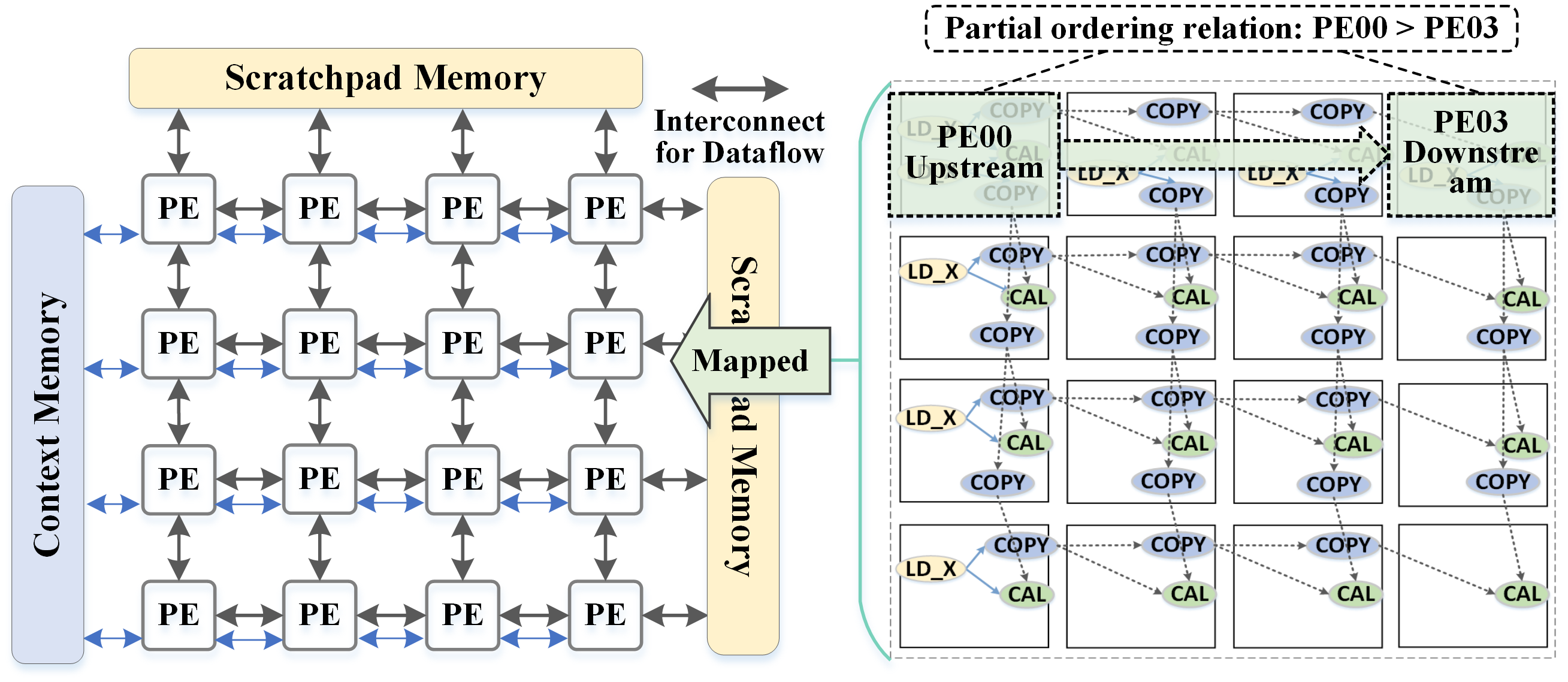}
    \vspace{-0.4cm}
    \captionsetup{font=small}
    \caption{Coarse-grained reconfigurable dataflow substrate mapped with GEMM DFG.}
    \label{fig:dfg_demo}
\vspace{-0.5cm}
\end{figure}

\begin{figure*}[t]
    \centering
    \hspace*{\fill}
    \begin{subfigure}{0.25\linewidth}
        \centering
        \includegraphics[width=\linewidth]{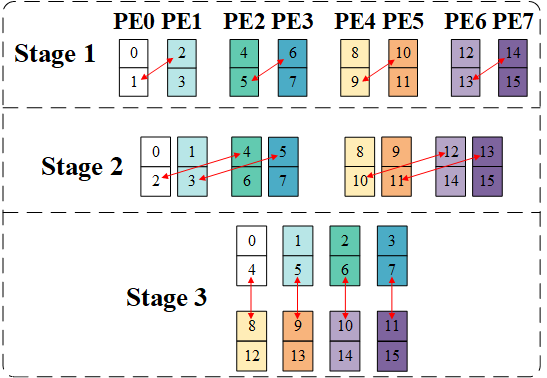}
        \captionsetup{font=small}
        \caption{A verification case of the butterfly swaps in 16 points.}
        \label{fig:swap_validity}
    \end{subfigure}
    \hfill
    \begin{subfigure}{0.39\linewidth}
    \includegraphics[width=\linewidth, height=0.46\linewidth]{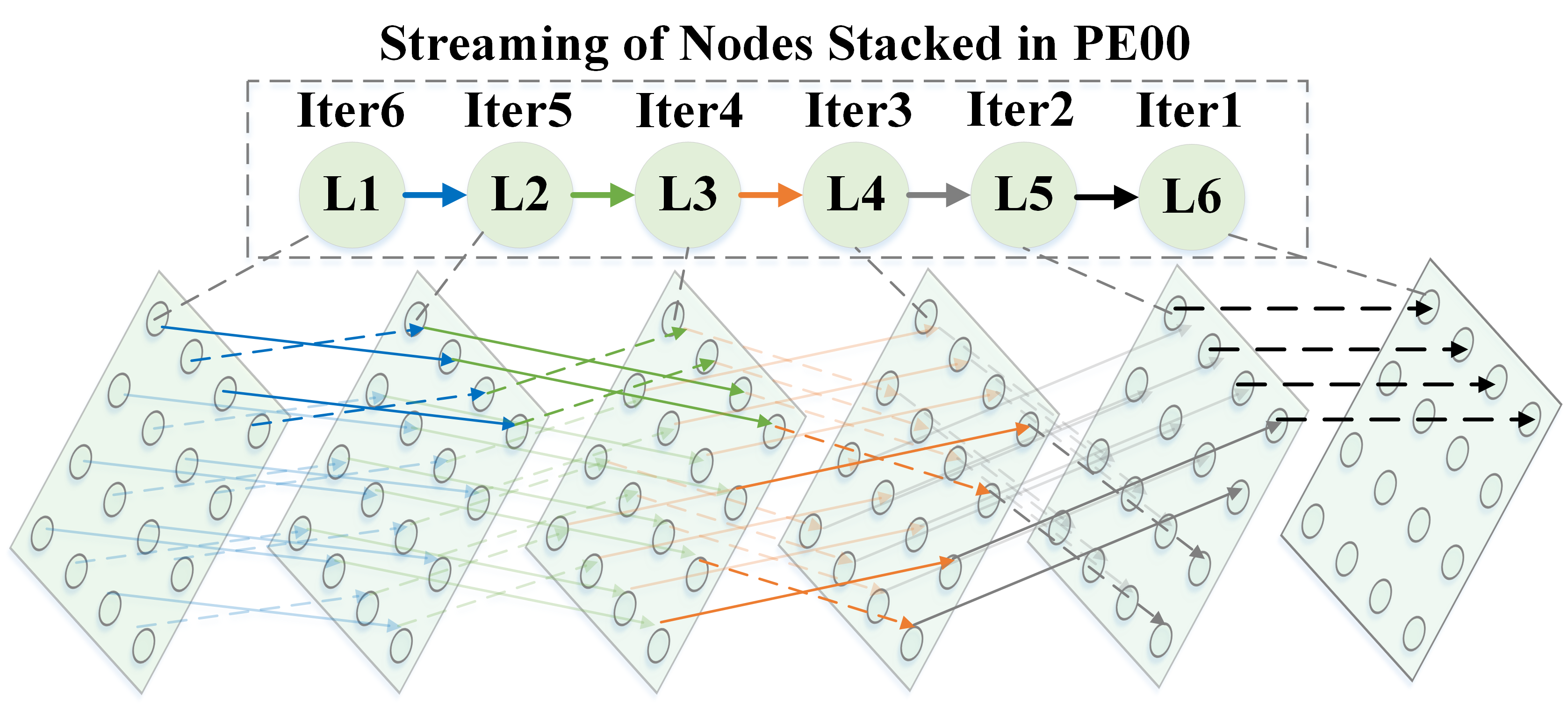}
    \vspace{-0.35cm}
    \captionsetup{font=small}
    \caption{Multilayer butterfly DFG execution with node streaming inside PE unit.}
    \label{fig:dataflow_graph}
    \end{subfigure}
    \hfill
    \begin{subfigure}{0.34\linewidth}
    \includegraphics[width=\linewidth, height=0.50\linewidth]{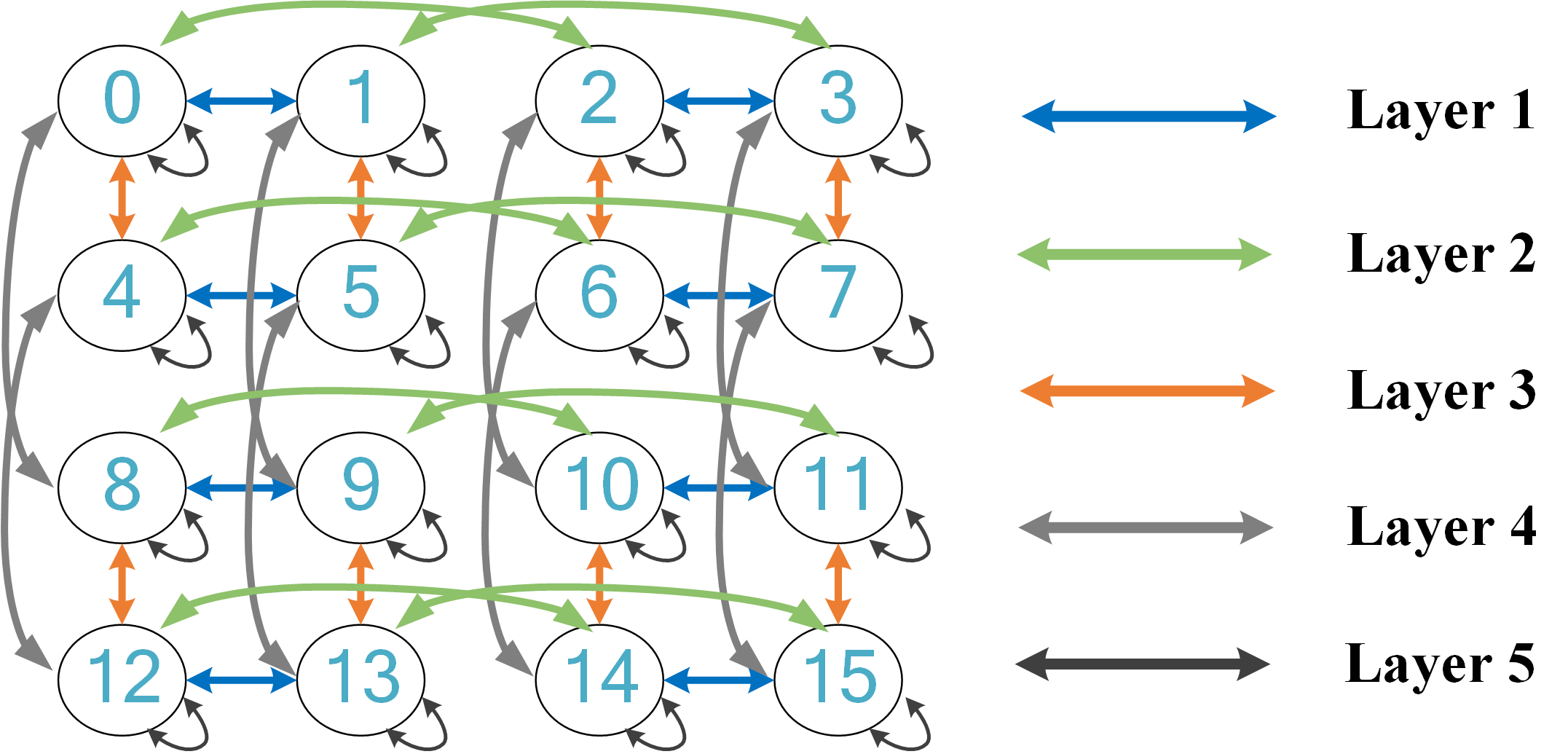}
    \vspace{-0.35cm}
    \captionsetup{font=small}
    \caption{Sufficient data flowing between PEs on mesh NoC.}
    \label{fig:butterfly_noc}
    \end{subfigure}
    \hspace*{\fill}
    \vspace{-0.35cm}
    \captionsetup{font=small}
    \caption{Multilayer dataflow orchestration exploiting data reuse on spatial PE array.}
    \label{fig:pe_flow_graph}
    \vspace{-0.45cm}
\end{figure*}

\subsection{The Substrate of Reconfigurable Dataflow Architecture}

Reconfigurable spatial architecture\cite{DySER} with data-driven execution shows superiority for its \emph{computational efficiency} and \emph{flexibility}. As shown in Fig.~\ref{fig:dfg_demo}, customized PE units are connected with a mesh-like network onto chip (NoC) to implement diverse data reuse. Dataflow graphs (DFG) compiled from workloads can be mapped on PE array with optimized strategies, which can flexiblely cover various operators in NNs \cite{plasticine}. Data dependence relationship between two nodes in DFG is explicitly \emph{partial ordering} - from upstream (PE00) to downstream (PE03). Input vector flows from scratchpad memory (SPM) into the PE array by \emph{LOAD} operations, and arrows indicate the data transfer of \emph{COPY} operations to satisfy the data dependence of calculation - \emph{CAL} inside PEs. PE units are simply designed with lightweight function units for the consideration of low energy consumption, while \emph{single instruction multiple data} (SIMD) is introduced to improve calculation parallelism for boosting computation performance.

\subsection{Butterfly Swap in Partial Ordering DFG}
 Fig.~\ref{fig:butterfly_before} depicts the dataflow of butterfly sparsity computation conducted through two matrices $B_i$. Each butterfly calculation is encapsulated as a single DFG node mapped onto a PE. These nodes require two input elements of vectors $X_i$ and the pre-stored static weights of matrices $B_i$, and they produce two output elements of vector $X_{i+1}$. Half of the output elements of a DFG node should be swapped with another node situated on the other PE. However, this mutual swap between DFG nodes in $B_i$'s multiplying is not in coordinate with the \emph{partial ordering relation} defined in DFG, because the \emph{dependence relation of upstream and downstream} is not explicit in PE00 and PE01.

To tackle this incoordination of partial ordering relation in the butterfly DFG, we extend the graph nodes into multiple layers, rearranging the butterfly swaps as Fig.~\ref{fig:butterfly_after}. In the first layer, two elements needed for inner calculations are fetched from SPM. Subsequently, each node retains half of the elements through a local transfer \emph{COPY\_I} inside PE unit, while the other elements flow to the node of the next layer via a remote transfer \emph{COPY\_T} at the sequential distances of 1, 2, 4, 8, and so forth. Fig.~\ref{fig:swap_validity} shows a simplified case of this flowing process. Vector elements from 0 to 15 are distributed among 8 PEs in order. In \emph{stage 1}, swaps are applied to elements 1 and 2, 5 and 6, ..., 13 and 14. \emph{In stage 3}, final swaps are performed on elements 4 and 8, 5 and 9, ..., 7 and 11. 

\section{Multilayer Dataflow Orchestration for Butterfly Sparsity} 
\vspace{-0.15cm}
\subsection{Mapping Multilayer Dataflow onto PE array}
The principle of mapping the multilayer DFG of butterfly matrices multiplication onto PE array is to ensure workload balancing as well as sufficient data reusing. A computation example of a 32-point vector on a $4\times4$ array is illustrated. As the 6-layer DFG shown in Fig.~\ref{fig:dataflow_graph}, graph nodes of different layer stages are distributed across the PE array in balance. Each PE accommodates one node per stage with a total of six. Colored arrows between nodes of different layers in Fig.~\ref{fig:dataflow_graph}, manifest as the arrows with the same color between PEs connected by the mesh NoC in Fig.~\ref{fig:butterfly_noc}. Therefore, this DFG mapping implements multilayer data flowing on the whole PE array with layer-level dependence between graph nodes, while sufficiently utilizing all the vertical and horizontal data paths of NoC in full throughput for efficient data reusing. 

Within each PE, the stacked nodes of all layers create a data-driven effect as streaming-like execution. For attention-based workloads, data dimensions of \emph{batch size} and \emph{head dimension} can provide data parallelism to pour adequate \emph{graph iterations} into the multilayer DFG in a pipelined way, which can boost the overall computation throughput within the PE array. It should be noted that under this array mapping method with quadratic combination distance (1,2,4,...), the graph node in a PE paired with another node at a distance greater than 16 for butterfly swap is wrapped back to the same PE (e.g., PE1 pairs with PE${17\%16}$=PE1), as the black arrows from layer 5 to layer 6 in Fig.~\ref{fig:dataflow_graph}. Therefore, butterfly swaps in later stages actually happen within the same PE, so further remote transfer is avoided. This careful orchestration of multilayer data dependence effectively avoids the massive concurrent data shuffling of different butterfly stages in shared SPM storage.
\vspace{-0.1cm}
\section{Parallelism Exploitation and Scalability} 
\subsection{Coarse-grained Streaming with Decoupled Units}

\begin{figure}[bp]
    \vspace{-0.55cm}
    \centering
    \includegraphics[width=\linewidth]{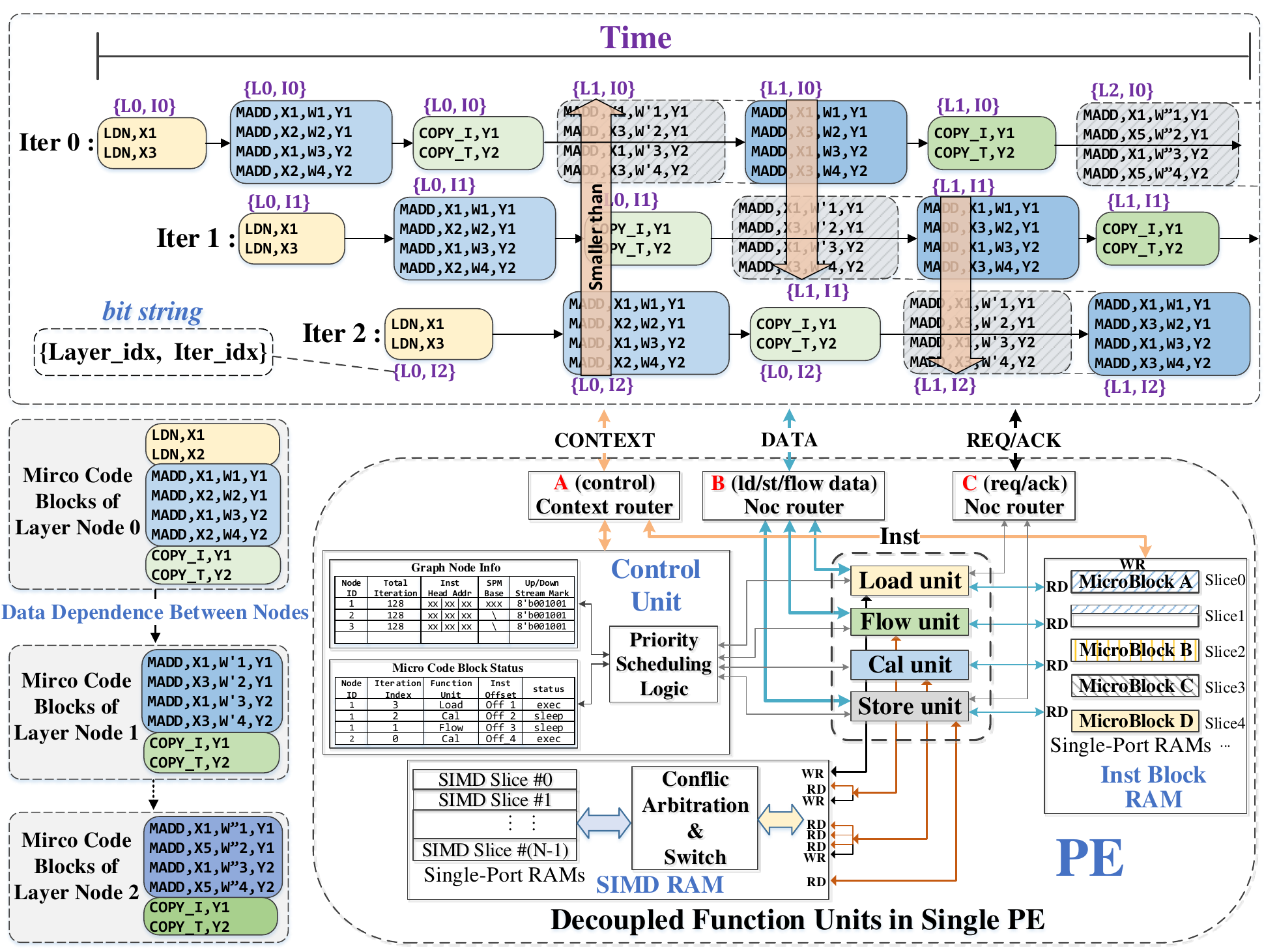}
    \vspace{-0.35cm}
    \captionsetup{font=small}
    \caption{Coarse-grained scheduling with decoupled function units.}
    \label{fig:block_scheduling}
\end{figure}

Unlike single-sequence scenarios found in \emph{signal processing}, the additional data dimensions \emph{(batch, head, hidden)} in attention workloads enable sufficient data parallelism and ensure nodes streaming with different iterations in a PE, as mentioned above. To manage this streaming parallelism, a \emph{coarse-grained block-level} scheduler is introduced in the PE's \emph{controlUnit}. Since tensor-based attention workloads have explicitly computational certainty (with few branch control), their instructions can be securely arranged into sequential \emph{Micro Code Blocks}, corresponding to the four function units as \emph{\{Load, Flow, Cal, Store\}}, as the colored code blocks shown in Fig.~\ref{fig:block_scheduling}.  

Each code block is scheduled, and it monopolizes the corresponding function unit until firing all the instructions. In Fig.~\ref{fig:block_scheduling}, the indexes of \emph{layer} and \emph{iteration} belonging to each block are concatenated as a priority bit string - \emph{\{Layer\_idx, Iter\_idx\}} to be judged by the block scheduling strategy in \emph{controlUnit}. The function unit scheduler compares the bit strings of all the running blocks and picks out the one with the smallest string number. This priority strategy ensures more DFG iterations to stream in for better \emph{exploiting block-level parallelism} as well as exhausting the utilization of all function units. Therefore, by scheduling coarse-grained code blocks instead of dispatching fine-grained instructions, four function units are decoupled from the instruction-level data dependence, and the arbitration logic inside \emph{controlUnit} can be simplified, which pursues the light-weight scheduling design philosophy for the consideration towards \textbf{\emph{better energy efficiency}}.

\begin{figure}[bp]
    \vspace{-0.6cm}
    \centering
    \includegraphics[width=\linewidth]{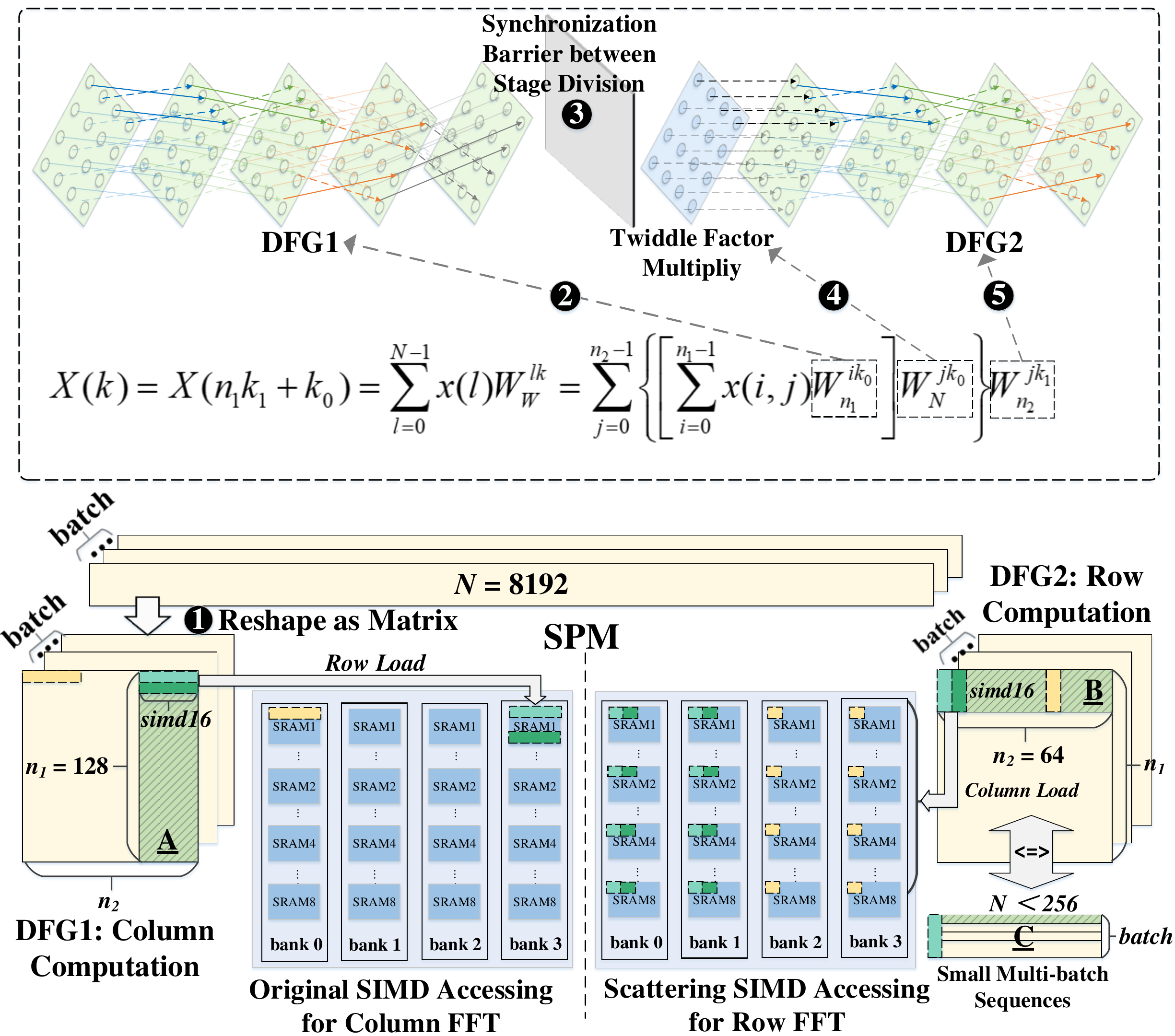}
    \vspace{-0.35cm}
    \captionsetup{font=small}
    \caption{Transpose-free 2D Cooley-Tukey factoring method using multi-line SPM for larger scale butterfly computation.}
    \label{fig:multi_stages}
\end{figure}

\subsection{Butterfly Sparsity of Long Vector Using Multi-stage DFGs}
In terms of the butterfly sparsity computation for longer vectors, on the basis of the \emph{Cooley\&Tukey algorithm}\cite{Cooley1965}, a multi-dimensional computation method with stage division is proposed to improve the data scalability of our design. The largest scale of the vector $X_i$ to be computed within a single multilayer DFG of BPMM or FFT is limited by the architecture resource in the dataflow substrate, like the SPM capacity or the register number in PE. In our design, the maximum DFG scale that can be mapped on the PE array is 256 for FFT (\emph{complex number}) and 512 for BPMM (\emph{real number}). 

Fig.~\ref{fig:multi_stages} illustrates the division process of an 8192-point vector example on our architecture. \ding{182} The 8192-point vector $X_{l=8192}$ is firstly reshaped as a $128\times64$ matrix $A_{r=128,c=64}$. \ding{183} Secondly, the first-stage butterfly computation executed as a 128-scale DFG1 is applied on the columns of $A_{r,c}$. \ding{184} Next, a synchronization barrier waiting for the finish of all iterations in DFG1 is set, ensuring all the intermediate data flowing back to the SPM under the resource constraint on PE array. \ding{185} Further, twiddle factors $\omega_N^{kn_2}$ are multiplied on $A_{r,c}$ in the form of an \emph{element-wise computation} layer (which is necessary for FFT but needless for BPMM). \ding{186} Finally, the third stage computation of a 64-scale DFG2 is executed on the rows of $A_{r,c}$. 

To compute an even longer vector with the data scale so large that can not fit into the SPM storage (e.g., a $64K$ vector whose sparsity weights occupy 8.4MB storage, while the SPM capacity in our design is 4MB.), the multi-stage reshaping method still works. For example, the $64K$ vector can be reshaped as a $256\times256$ matrix. The two-stage 256-scale DFG1 and DFG2, each of which covers half of the butterfly sparsity weights, can be launched in two times on the PE array, with their weights or twiddle factors swapping between SPM storage and DDR.

\begin{figure}[t]
    \centering
    \includegraphics[width=\linewidth]{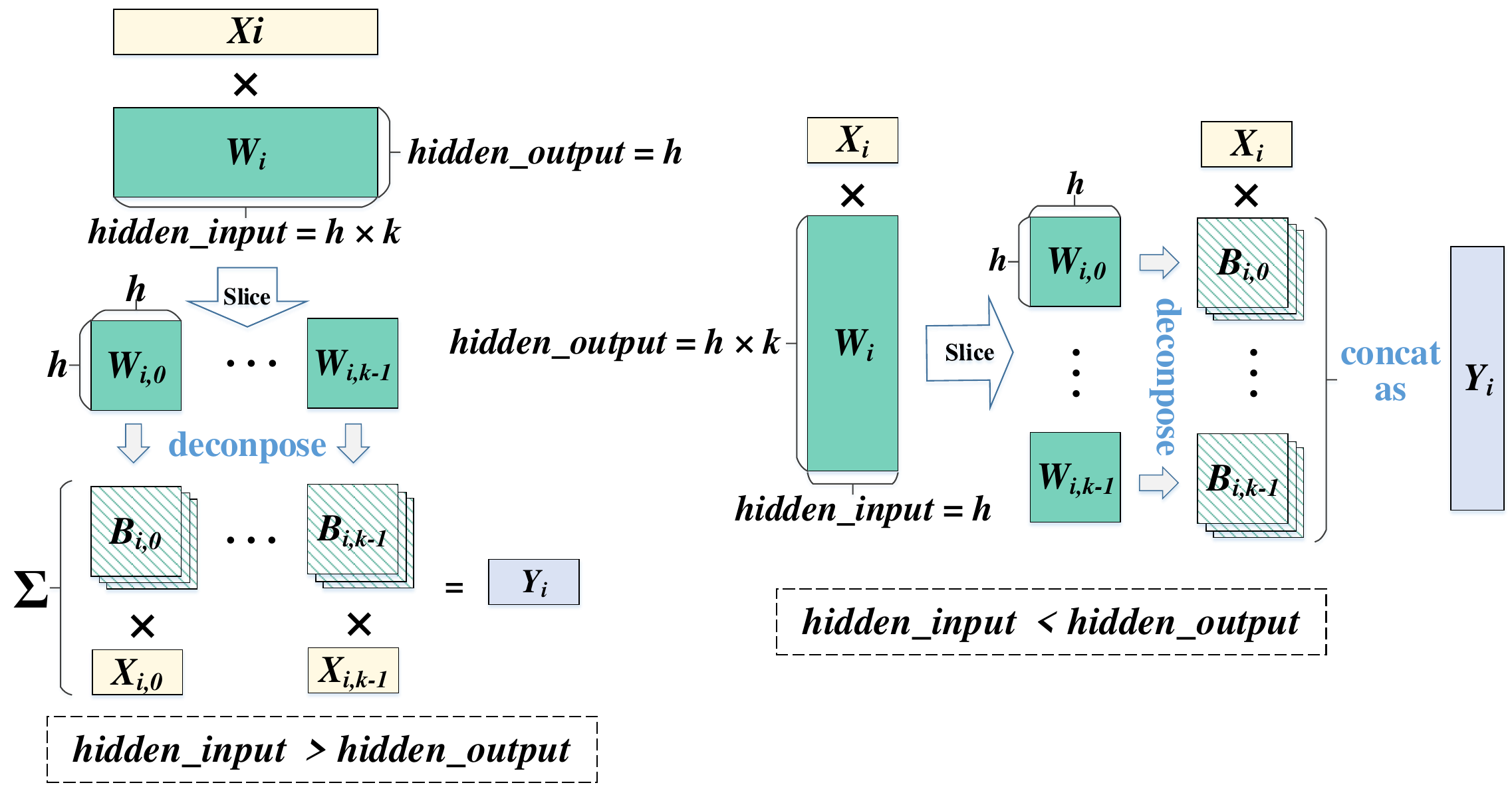}
    \vspace{-0.3cm}
    \captionsetup{font=small}
    \caption{Weight matrix slicing with unequal input and output hidden size in BPMM computation.}
    \label{fig:weight_slice}
\vspace{-0.5cm}
\end{figure}

In terms of BPMM, the linear weight matrices of ${Q,K,V}$ or \emph{FFN} may have different input and output hidden sizes, so they should be sliced for applying butterfly sparsity decomposition. As shown in Fig.~\ref{fig:weight_slice}, if the hidden size of the input vector $X_i$ is larger than the size of the output vector $Y_i$, both the weight matrix $W_i$ and the $X_i$ will be sliced into multiple pieces. Each sliced $W_{i,j}$ makes decomposition as the butterfly matrices $B_{i,j}$ and will be multiplied by the vector piece $X_{i,j}$. All these multiplied products will be summed up as the final result $Y_i$. Otherwise, in the smaller situation, the product piece of $Y_{i,j}$ (produced by the decomposed $B_{i,j}$ and the shorter-sized $X_i$) will be concatenated as the longer-sized output $Y_i$.

\begin{table*}[htp]
    \vspace{-0.15cm}
    \centering
    \caption{Platform Comparison and Benchmark.}
    \vspace{-0.05cm}
    \renewcommand{\arraystretch}{1.3}
\begin{tabular}{|l|cccc|}
\hline
\rowcolor[HTML]{DDD9D9} 
Platform Comparison & \multicolumn{1}{c|}{\cellcolor[HTML]{DDD9D9}Jetson Nano} & \multicolumn{1}{c|}{\cellcolor[HTML]{DDD9D9}\textbf{SOTA Butterfly Acc \cite{Adaptable}}} & \multicolumn{1}{c|}{\cellcolor[HTML]{DDD9D9}\textbf{Jetson Xavier NX}} & \textbf{Multilayer Dataflow} \\ \hline
Frequency & \multicolumn{1}{c|}{921MHz} & \multicolumn{1}{c|}{200MHz (FPGA)} & \multicolumn{1}{c|}{1.1GHz} & 1GHz \\ \hline
Performance (fp16) & \multicolumn{1}{c|}{471.6 GFLOPs} & \multicolumn{1}{c|}{204.8 GFLOPS (512 MACs)} & \multicolumn{1}{c|}{1.69 TFLOPS (CUDA),} & 1.02 TFLOPS (512 MACs) \\
 & \multicolumn{1}{l|}{} & \multicolumn{1}{l|}{} & \multicolumn{1}{l|}{11 TFLOPS (Tensor)} & \multicolumn{1}{l|}{256 GFLOPS (128 MACs)} \\ \hline
Bandwidth Supply & \multicolumn{1}{c|}{25.6 GB/s} & \multicolumn{1}{c|}{21.3 GB/s} & \multicolumn{1}{c|}{59.71 GB/s} & 25.6x2 GB/s \\ \hline
Technology & \multicolumn{1}{c|}{20nm} & \multicolumn{1}{c|}{28nm} & \multicolumn{1}{c|}{12nm} & 12nm \\ \hline
Power Consumption & \multicolumn{1}{c|}{10W} & \multicolumn{1}{c|}{11.355W} & \multicolumn{1}{c|}{15W} & 6.95W (DC Synthesized) \\ \hline
        \rowcolor[HTML]{D9D4D4} 
        \cellcolor[HTML]{D9D4D4}Benchmark & \multicolumn{4}{c|}{\cellcolor[HTML]{D9D4D4}Platform Selected for Comparison}    \\ \hline   VIT and BERT (BPMM and FFT)    & \multicolumn{1}{c|}{}   & \multicolumn{1}{c|}{}  & \multicolumn{1}{c|}{\textcolor{black}{\ding{52}}}   & \textcolor{black}{\ding{52}}     \\ \hline
        FABNet-Base Transformer           & \multicolumn{1}{c|}{\textcolor{black}{\textbf{\ding{52}}}}   & \multicolumn{1}{c|}{\textcolor{black}{\ding{52}}}  & \multicolumn{1}{c|}{}   & \textcolor{black}{\ding{52}}    \\ \hline
        One-layer Vanilla Transformer    & \multicolumn{1}{l|}{}   & \multicolumn{1}{c|}{\textcolor{black}{\ding{52}}}   & \multicolumn{1}{l|}{}  & \textcolor{black}{\ding{52}}         \\ \hline
        \end{tabular}
    \label{platforms}
    \vspace{-0.5cm}
\end{table*}


\subsection{Transpose-free SIMD Parallelism with Multi-line SPM}
To utilize the adequate data parallelism provided by the data dimension of attention layers, the calculation units in our PEs are designed with SIMD to boost the overall performance. However, the mentioned multi-stage Cooley\&Tukey method introduces SIMD alignment inconsistency in column and row stages. As the reshaped matrix shown in Fig.~\ref{fig:multi_stages}, the DFG1 computes in columns and accesses SIMD in rows, while the DFG2 computes in rows and accesses SIMD in columns. To unify these two accessing patterns while avoiding redundant matrix transpose operation, a multi-line SPM design is introduced. Firstly, the entry width of SRAMs in SPM is set to SIMD16, matching the calculation width in the PE. And the address mapping of SRAM entries is interleaved among 4 banks in SPM unit, to utilize bank-level parallelism\cite{bank-level}. Secondly, each bank is architected with eight lines, so two banks can be accessed in parallel to give out SIMD16 from all lines. In $A_{r,c}$, elements of SIMD16 in rows can be accessed entry by entry from single SRAM, while elements in columns are scattered among 16 lines in banks in the form of \{\emph{element0$\mapsto$bank0\_line0, e1$\mapsto$b0\_l1, e2$\mapsto$b0\_l2, ..., e8$\mapsto$b1\_l0, e9$\mapsto$b1\_l1, ..., e15$\mapsto$b1\_l7}\}.

Under this data organization in SPM, SIMD alignments for butterfly sparsity in various data scales are achieved, as shown on the right of Fig.~\ref{fig:multi_stages}. \textbf{\underline{A}.} The first alignment of the column butterfly in DFG1 is trivially satisfied by using a normal \emph{row-wise} load operation in SRAM entries. \textbf{\underline{B}.} The second alignment of the row butterfly in DFG2 requires a particular \emph{column-wise} load to gather 16 elements from lines. \textbf{\underline{C}.} Moreover, in terms of the vectors in small scales (e.g., shorter than 256 and no need for matrixing) but in multiple batches, their data layout is actually identical to the \emph{column-wise} accessing in DFG2. These short vectors can be scattered among SPM lines so that the \emph{batch dimension} can be aligned to the SIMD lanes. This multi-line SPM architecture is friendly with the continuous burst data transfer of DMA and has high bandwidth utilization of DDR.

\vspace{-0.1cm}
\section{Experiments}
\subsection{Experiment Settings}
\textbf{Methodology:} We develop a parallel cycle-accurate simulator based on PDES\cite{PDES} framework for the performance statistic of running butterfly attentions on the dataflow substrate. An RTL fabric of the dataflow substrate equipped with coarse-grained scheduling is implemented and synthesized with Synopsys Design Compiler and 12nm TSMC standard library at 1 GHz frequency, for the evaluation of hardware resource and power consumption. The cycle error rate of the simulator is calibrated within 7\% over the RTL design. All butterfly sparsity kernels in various data scales are assembled with an optimized DFG template to generate the dependence relations and the micro codes of the graph blocks mapped onto the PE array.

\textbf{Baselines and benchmark:} \textbf{\underline{A}.} We first make performance comparison over NVIDIA - Jetson Xavier NX equipped with powerful tensor cores to evaluate the computation efficiency of butterfly sparsity running on multilayer dataflow design. We implement the butterfly kernels of FFT replacing $softmax(qk)*v$ as well as BPMM replacing \emph{linear layers} in both VIT and BERT as the first benchmark. \textbf{\underline{B}.} Next, we select the work\cite{Adaptable} as the state-of-the-art (SOTA) baseline to evaluate the performance and utilization improvement of our architectural superiority in attention computations, taking the original butterfly sparsity network - \emph{FABNet-Base Transformer} in \cite{Adaptable} as the second benchmark, while using NVIDIA Jeston Nano as the \underline{normalized object} for speedup ratio comparisons. Experiment results of the GPU platform are obtained from the NVIDIA profiling tool - Nsight Compute. Detailed parameter and configurations are listed in Table.~\ref{platforms}. The butterfly kernels implemented on GPU are based on the \emph{cuFFT library}, and all the attention workloads focus on \emph{forward inference}. 

\vspace{-0.1cm}
\subsection{Model Accuracy with Butterfly Sparsity}
\begin{figure}[bp]
    \vspace{-0.5cm}
    \centering
    \includegraphics[width=\linewidth]{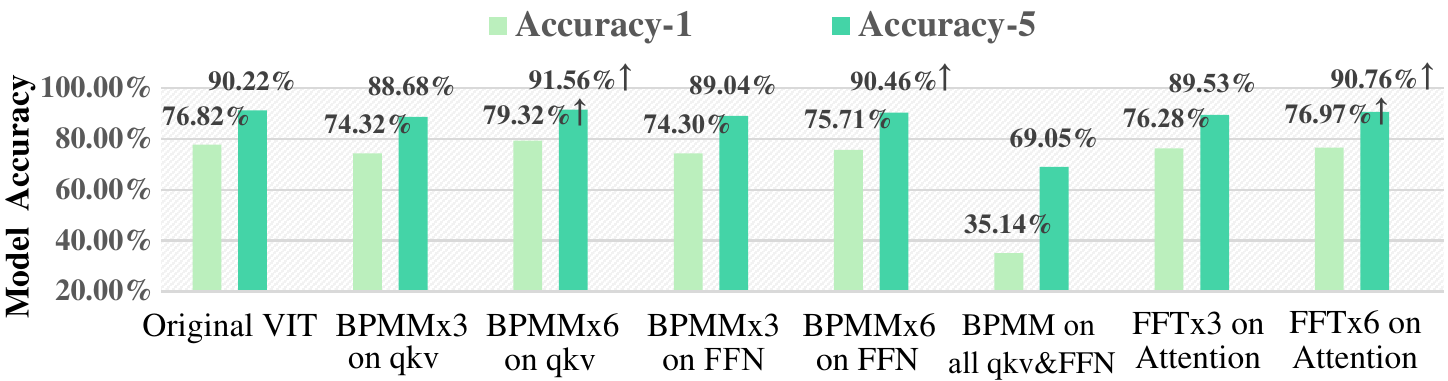}
    \vspace{-0.35cm}
    \captionsetup{font=small}
    \caption{VIT Model trained with butterfly sparsity after 1000 epochs.}
    \label{fig:accuracy}
\end{figure}

Accuracy estimations of model VIT with BPMM on linear weights $q,k,v$ and \emph{FFN} as well as FFT on \emph{Attention matrices} are shown in Fig.~\ref{fig:accuracy}. The overall accuracies of top-1 and top-5 have limited loss of less than 2.6\% over the original VIT network, except for the case with all linear layers in $q,k,v$, and \emph{FFN} replaced by BPMMs. An unexpected result is that the case of $q,k,v$ in 6 attention layers replaced by BPMMs has accuracy improvements from 76.82\% and 90.22\% to 79.32\% and 91.56\%, after training with 1000 epochs. Also there is slight accuracy improvement (76.97\% and 90.76\%) in the case replaced with 6 FFT on attention layers. These results highlight the weight compression effect by introducing butterfly sparsity to speed up model convergence, and there can be a tradeoff between performance and accuracy in layer-level prunning. As shown in Table.~\ref{table:LLaMa}. Applying comression on a certain amount of layers do not greatly affect accuracy. About the generative model LLaMa, the frontal layers from 1 to 12 of it are evaluated to have smaller STDs instead, compared with the remaining 20 layers. We apply consecutive 3-layer Fourier compressions on different layer segments with fine-tuning and evaluate the model metrics. The results basically consist with BERT, implying that it's effective by using STD indication to help for selecting model layers and scaling truncation.

\begin{table}[tp]
\vspace{-0.2cm}
\captionsetup{font=footnotesize}
\caption{Accuracy of bert and LLaMa-DeepSeekCoder using compression.}
\vspace{-0.1cm}
\label{table:LLaMa}
\captionsetup{font=footnotesize}
\resizebox{0.495\textwidth}{!}{
\centering
\renewcommand{\arraystretch}{1.25}
\begin{tabular}{l|c|ccccc}
    \hline
    \textbf{BERT, N=512} & \textbf{Original} & \multicolumn{1}{l}{\textbf{1 Layer}} & \multicolumn{1}{l}{\textbf{3 Layers}} & \multicolumn{1}{l}{\textbf{6 Layers}} & \multicolumn{1}{l}{\textbf{9 Layers}} & \textbf{12 Layers} \\ \hline
    \textbf{F1} & 86.93 & \textbf{87.04} & 86.52 & 86.6 & 85.77 & 84.93 \\
    \textbf{Exact} & 78.37 & 78.24 & 77.81 & 77.87 & 76.96 & 75.92 \\ \hline
    \textbf{LLaMa, N=4K} & \textbf{Original} & \textbf{L1-L3} & \textbf{L3-L5} & \textbf{L5-L7} & \textbf{L7-L9} & \textbf{L9-L12} \\ \hline
    \textbf{Similarity} & \multicolumn{1}{c|}{76.86} & 75.32 & 74.45 & 75.33 & \textbf{75.92} & 75.30 \\
    \textbf{Exact Match} & \multicolumn{1}{c|}{42.18} & 39.42 & 38.56 & 39.07 & \textbf{40.62} & 39.41 \\
    \textbf{Avg STDs} & \multicolumn{1}{c|}{\textbf{}} & 1.22 & 2.60 & 2.27 & \textbf{0.95} & 1.91 \\ \hline
    \end{tabular}
    }
    \vspace{-0.5cm}
\end{table}

\vspace{-0.05cm}
\subsection{Accessing Inefficiency Alleviation with Dataflow}

\begin{figure}[ht]
    \centering
    \includegraphics[width=\linewidth]{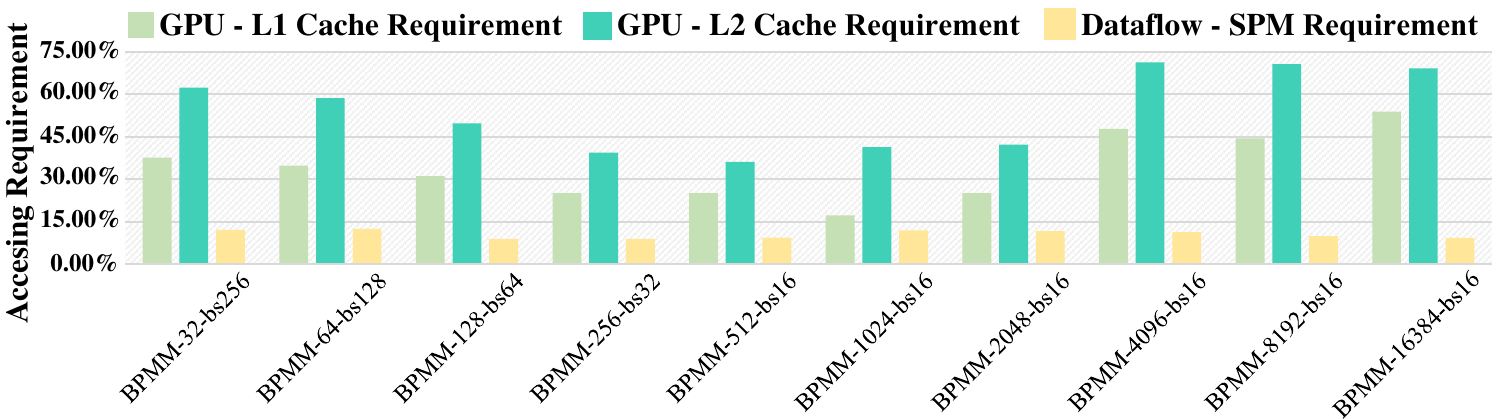}
    \vspace{-0.4cm}
    \captionsetup{font=small}
    \caption{Data accessing requirement percentages of cache and SPM for the GPU and multilayer dataflow.}
    \label{fig:access_throughput}
\vspace{-0.2cm}
\end{figure}

Data-accessing throughput percentages of the cache in Jetson Xavier NX and the SPM in our multilayer dataflow are shown in Fig.~\ref{fig:access_throughput}. The accessing requirement of L1 cache in NX is over 20\% (up to 53.80\%), while  the requirement is over 40\% (up to 71.19\%) for L2 cache. Both of them increase with the sequence scale at greater than 512 because of the more frequent data shuffling with larger accessing strides in the later butterfly stages on GPU. However, thanks to the multilayer DFG orchestration bringing sufficient data reuse on PE array and the multi-bank SPM design avoiding extra transpose operation, the overall accessing requirement from SPM is compressed to below 12.48\%, which shows the alleviation effect of our dataflow method on the accessing bottleneck in butterfly sparsity computation.
\vspace{-0.1cm}
\subsection{Effectiveness of Coarse-grained Decoupling}
\begin{figure}[tp]
    \centering
    \hspace*{\fill}
    \begin{subfigure}{0.98\linewidth}
    \includegraphics[width=\linewidth]{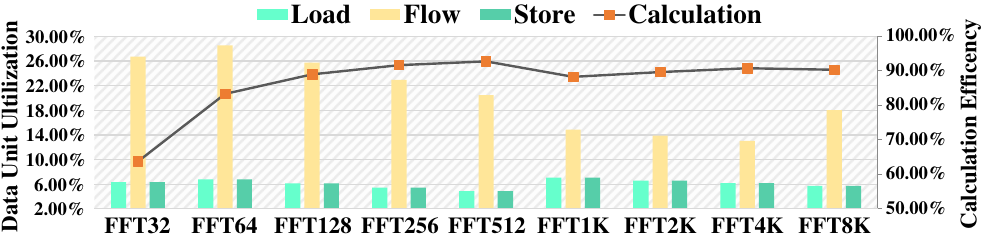}
    \vspace{-0.4cm}
    \captionsetup{font=footnotesize}
    \caption{Decoupled units utilization of FFT on attention.}
    \label{fig:decouple_fft}
    \vspace{+0.2cm}
    \end{subfigure}
    \hfill
    \begin{subfigure}{0.98\linewidth}
    \includegraphics[width=\linewidth]{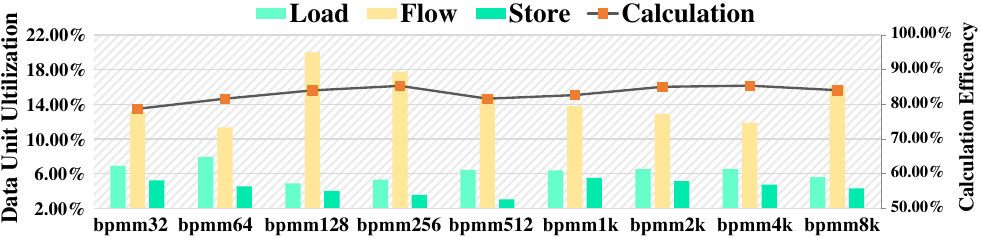}
    \vspace{-0.4cm}
    \captionsetup{font=footnotesize}
    \caption{Decoupled units utilization of BPMM on linear layers.}
    \label{fig:decouple_bpmm}
    \end{subfigure}
    \hspace*{\fill}
    \captionsetup{font=small}
    \vspace{-0.3cm}
    \caption{The effectiveness of the coarse-grained decoupled units design for butterfly sparsity computation.}
    \label{fig:decouple_result}
    \vspace{-0.5cm}
\end{figure}

The utilization of four decoupled units is illustrated in Fig.~\ref{fig:decouple_result}. Although the unit executions are scheduled by code blocks in a coarse-grained manner, the multilayer DFG mapped onto the PE array provides adequate block-level parallelism to hide the accessing latency from SPM and the switching latency between iterations. Under this scheduling, the utilization of \emph{calculation units} is over 64\% for all butterfly kernels and above 89\% for FFT in large scales. For the other data units, however, the utilization of \emph{Load} is less than 6\% for FFT and 8\% for BPMM, which indicates the sufficient data reuse of our orchestration method that avoids remote data fetching. What differentiates BPMM from FFT kernels is the utilization of lower \emph{Flow} and higher \emph{Load} in the same sequence scales. The real-valued BPMM has lower arithmetic density and needs more element loading of input vectors and weights, while the complex-multiplication FFT requires twice \emph{Flow} operations (20.45\% on average) to swap the real and imaginary parts of the intermediate $X_{i}$. This overall calculation efficiency among BPMM and FFT in different scales demonstrates the coarse-grained scheduling effectiveness in butterfly dataflow execution.

\subsection{Scalability with Different Stage Division}
\begin{figure}[t]
    \centering
    \includegraphics[width=\linewidth]{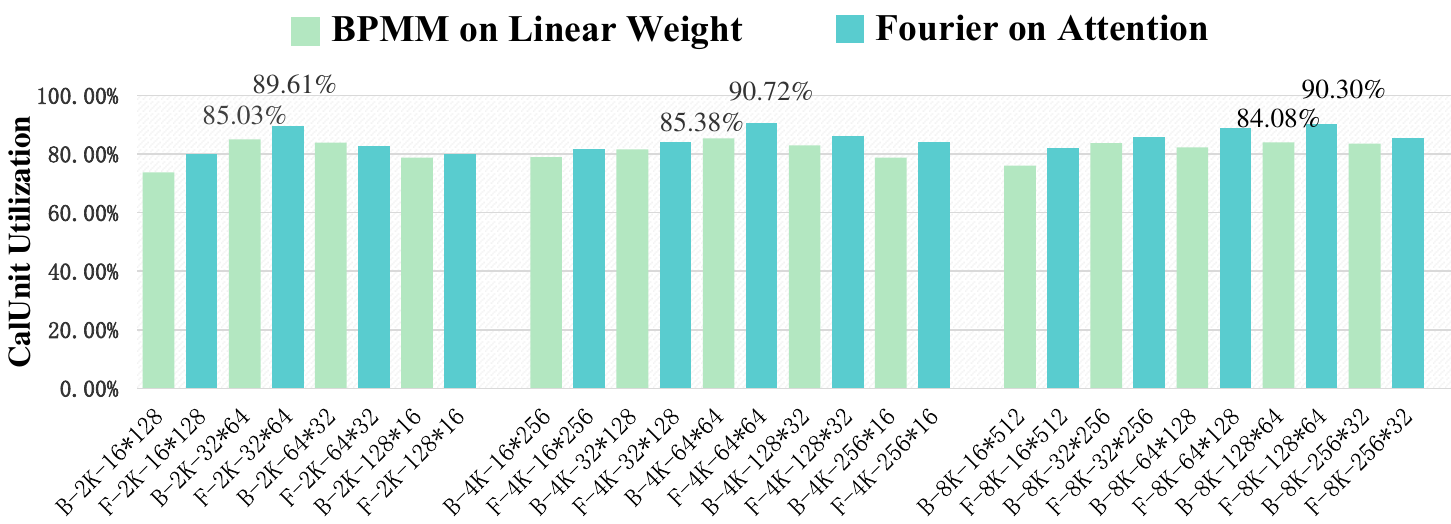}
    \vspace{-0.35cm}
    \captionsetup{font=small}
    \caption{CalUnit utilization in different sequence divisions.}
    \label{fig:balance_division}
\end{figure}

For long vectors, there are different stage division schemes to scale down the DFGs of butterfly kernels on PE array, as evaluated in Fig.~\ref{fig:balance_division}. The best divisions respectively for BPMM-2k, 4k and 8k, with the highest \emph{calUnit} utilization are 32*64 (85.03\%), 64*64 (85.38\%), 128*64 (84.08\%), which indicates the balancing tendency for the better performance in BPMM. This balanced division is also appropriate for FFT in large scales. These results make sense that since the multilayer DFG execution implements data reuse by data flowing on the mesh NoC, there will be inadequate reuse for the shallow stage (e.g., 16 points) to overlap the data fetching latency in an unbalanced division, in which a marginal effect may happen to the other deep stage (e.g., 256 points).

\subsection{Attention Performance over GPU}
\begin{figure}[tp]
    \vspace{-0.2cm}
    \centering
    \includegraphics[width=\linewidth]{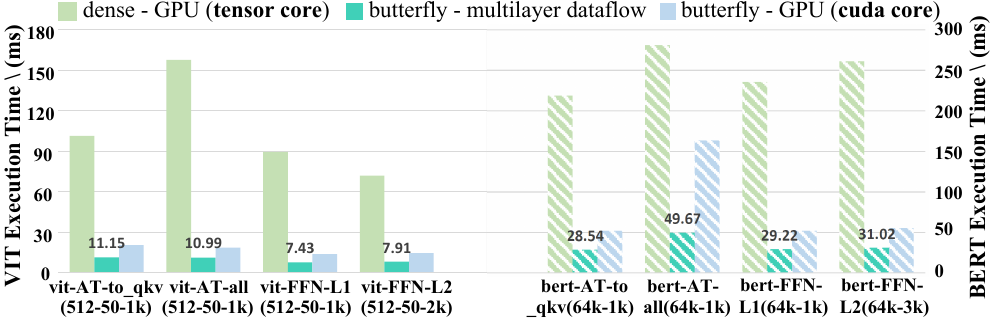}
    \captionsetup{font=small}
    \vspace{-0.35cm}
    \captionsetup{font=small}
    \caption{Execution time comparison over Jetson Xavier NX with tensor (dense) and cuda (butterfly-sparse) cores.}
    \label{fig:attention_time}
    \vspace{-0.4cm}
\end{figure}  

Execution times of attention kernels on Jetson Xavier NX and dataflow design are illustrated in Fig.~\ref{fig:attention_time}. The \emph{dense} prefix represents the attention kernel running without any sparsity on NX GPU. The \emph{AT-to\_qkv} and \emph{FFN-Lx} are the linear layers that are sparse with BPMM, while \emph{AT-all} is the whole attention layer that is sparse with 2D-FFT.
In comparison to Jetson Xavier NX with powerful \textbf{\emph{tensor cores}} (11 TFLOPs) running dense attention kernels, our design running butterfly kernels achieves a speedup of up to $14.34\times$ ($11.13\times$ on average) for VIT kernels and up to $8.42\times$ ($7.45\times$ on average) for BERT. The speedup effects on the \emph{AT-all} kernels with 2D-FFT sparsity are more obvious than others with BPMM, because the FFT replacement on attention layers has higher sparsity than BPMM while our design has equally excellent calculation efficiency on both FFT (over 89\%) and BPMM. 

In comparison to Jetson Xavier NX with pure \textbf{\emph{cuda cores}} (1.69 TFLOPs) also running butterfly kernels, our design (1.02 TFLOPs) achieves a speedup of up to $1.82\times$ ($1.78\times$ on average) for VIT under the $1.67\times$ peak performance gap in between. The average speedup is $1.97\times$ for BERT of larger 64k sequence scales, and the heaviest attention kernel - \emph{BERT-AT-all (64K sequences, 1K hidden)} with FFT sparsity has the maximum speedup of $3.30\times$ on the multilayer dataflow. This kernel is executed in 3-stage FFT of one-time 1K-point FFT on \emph{hidden} and two-time 256-point FFT on \emph{sequence}, on the basis of the scalable stage division method. This division ensures adequate point scale for sufficient flowing on PE array in each stage, while avoiding frequent data shuffling with huge strides in SPM. The speedup results show the excellent computational efficiency with dataflow orchestration for butterfly sparsity.

\subsection{Hardware Overhead and Energy Efficiency}
\begin{figure}[bp]
    \vspace{-0.5cm}
    \includegraphics[width=\linewidth]{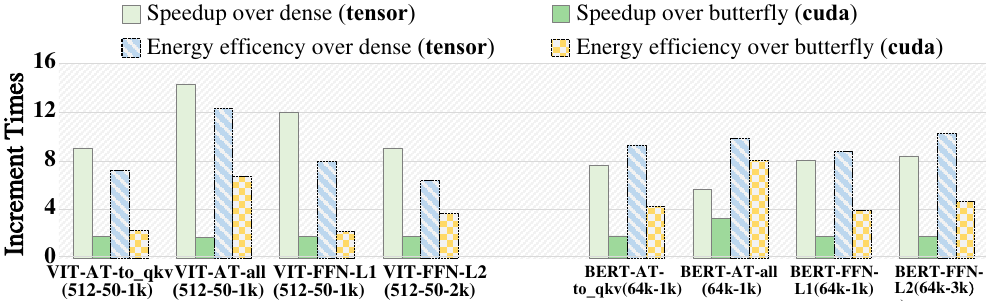}
    \vspace{-0.3cm}
    \captionsetup{font=footnotesize}
    \caption{Speedup and energy efficiency over GPU with tensor/cuda core.}
    \label{fig:speedup_energy}
\end{figure}

The hardware overhead of our dataflow design is shown in Table.~\ref{table:HardOverhead}. The PE unit takes up a synthesized area of 0.985$mm^2$ at 12nm process, and the overall power consumption of the 16 PE array evaluated by DC is 6.95W. Under this hardware configuration, energy efficiency comparison with Jetson Xavier NX is shown in Fig.~\ref{fig:speedup_energy}. Dataflow design achieves energy efficiency increment from $6.38\times$ to $12.32\times$ for dense kernels running with tensor cores, and from $2.17\times$ to $8.06\times$ over the butterfly kernels with CUDA cores. The attention kernel of FFT with higher arithmetic density acquires higher energy efficiency on our design. This efficiency advancement relies highly on the superb data reuse on PE array and the lightweight coarse-grained scheduling design inside PE.

\begin{table}[t]
    \captionsetup{font=small}
    \caption{Synthesized area and power consumption of PE unit.}
    \captionsetup{font=normalsize}
    \vspace{-0.1cm}
    \label{table:HardOverhead}
    \resizebox{\linewidth}{!}{
    \renewcommand{\arraystretch}{1.1}
    \begin{tabular}{l|l|l|l}
    \hline
    \textbf{Name}             & \textbf{Cell Area($mm^2$)} & \textbf{ActivePower(mW)} & \textbf{ActivePower(\%)} \\ \hline
    ContextRouter    & 0.018             & 6.37            & 1.47            \\
    DataRouter       & 0.108             & 62.21           & 14.31           \\
    ControlUnit       & 0.002             & 2.58          & 0.59            \\
    InstBlocks       & 0.039             & 9.23            & 2.12            \\
    SIMD RAM     & 0.106             & 32.13           & 7.39            \\
    FuncUnits (SIMD32)  & 0.316             & 322.16          & 74.11           \\
    Total (single PE) & 0.985             & 434.68 (6.95W for PE16 array)        & 100             \\ \hline
    \end{tabular}
    }
    \vspace{-0.3cm}
  \end{table}

\begin{figure}[bp]
    \vspace{-0.3cm}
    \includegraphics[width=\linewidth]{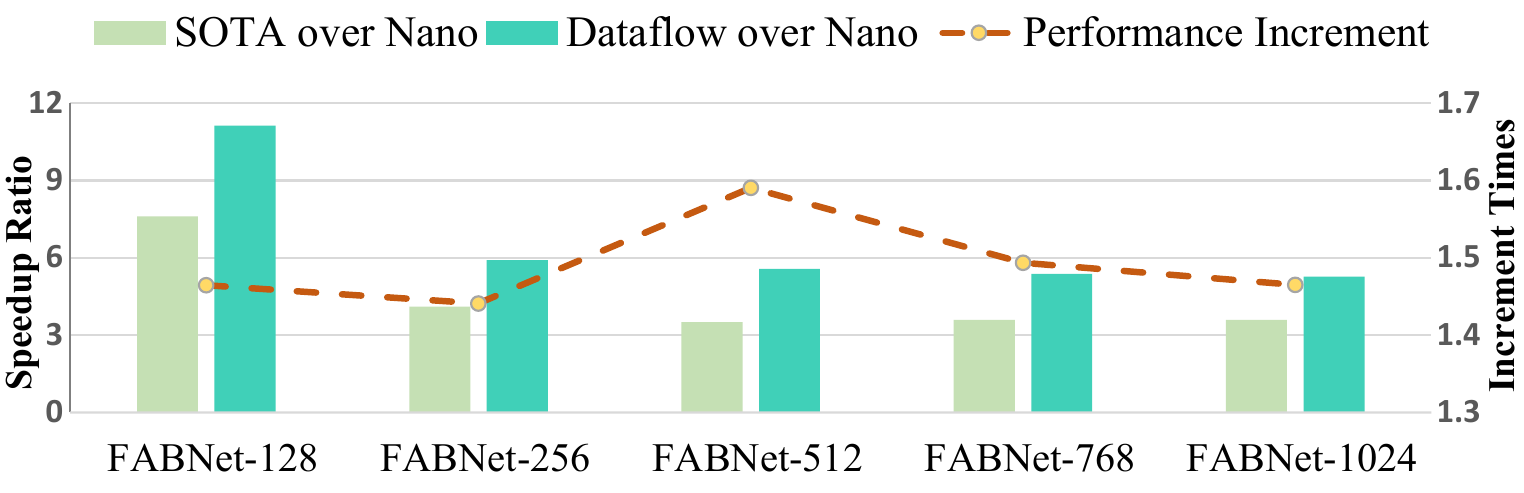}
    \vspace{-0.45cm}
    \captionsetup{font=footnotesize}
    \caption{Speedup comparisons over SOTA butterfly sparsity accelerator with the normalized object - Jetson Nano.}
    \label{fig:speedup_sota}
\end{figure}

\vspace{-0.1cm}
\subsection{Comparison with SOTA Works}
Since the SOTA butterfly accelerator is evaluated on FPGA platform with limited performance (204.8 GFLOPs) and bandwidth supply (21.3 GB/s), we scale down the number of MACs from 512 to 128, and halve the DDR channel on our design to keep fair performance comparison while taking Jetson Nano as the normalized speedup object. We select the \emph{FABNet transformer} \cite{Adaptable} as the workload which consists of 2D-FFT attention layers and BPMM FFN layers in sequence scales of 128, 256, 512 and 1K. As shown in Fig.~\ref{fig:speedup_sota}, our dataflow design has speedup ratios from $5.27\times$ to $11.13\times$ in comparison to the speedups from $3.5\times$ to $7.1\times$ of the SOTA butterfly accelerator, gaining a performance increment from $1.44\times$ to $1.59\times$ under the slight peak performance gap (256 versus 204.8 GFLOPs). The highest increment goes to the FABNet-512 whose working set size just fills up the SPM capacity (4MB) in our design scale, so all butterfly stages are executed in place, while extra DMA transfer from DDR is avoided and data reuse is exploited at most within PE array and SPM.

\begin{table}[t]
    \captionsetup{font=small}
    \caption{Latency and energy comparisons over other architectures.}
    \vspace{-0.1cm}
    \label{table:sota_e2e}
    \resizebox{\linewidth}{!}{
    \renewcommand{\arraystretch}{1.1}
    \begin{tabular}{l|cc|c|c}
        \hline
        \textbf{Accelerators} & \multicolumn{1}{c|}{\textbf{SpAtten}\cite{SpAtten}} & \textbf{DOTA}\cite{dota} & \textbf{SOTA Acc} & \textbf{Our work} \\ \hline
        \textbf{Technology} & \multicolumn{1}{c|}{ASIC (40nm)} & ASIC (22nm) & FPGA (28nm) & ASIC (12nm) \\ \hline
        \textbf{Frequency} & \multicolumn{2}{c|}{1 GHz} & 200 MHz & 1 GHz \\ \hline
        \textbf{Number of MACs} & \multicolumn{2}{c|}{128} & 640 & 128 \\ \hline
        \textbf{Latency (ms)} & \multicolumn{1}{c|}{48.8} & 34.1 & 2.4 & 2.06 \\ \hline
        \textbf{Throughput(Pred./S)} & \multicolumn{1}{c|}{20.49} & 29.32 & 416.66 & 485.43 \\ \hline
        \textbf{Power (W)} & \multicolumn{1}{c|}{1.06} & 0.858 & 11.355 & 3.94 (SIMD8 PE16) \\ \hline
        \textbf{Energy Eff.(Pred./J)} & \multicolumn{1}{c|}{19.33} & 34.18 & 36.69 & 123.21 \\ \hline
        \end{tabular}
        }
        \vspace{-0.4cm}
        \label{table:sota_e2e}
    \end{table}

We make overall end-to-end performance and energy efficiency comparison over SOTA and other accelerators that are designed towards \textbf{other dynamic sparsity variants}, under the premise of the same peak performance. Part of the evaluation results are quoted from the work \cite{Adaptable}. We select \emph{LAR-Image} dataset \cite{LRA-Image} on a one-layer vanilla transformer\cite{BERT} (1K-sequence and 1K-hidden) which is applied butterfly sparsity with 2D-FFT on attention matrix and BPMMs on two-layer FFN, as our benchmark. Input sequences are supplied in batch-256 and streamed in one-by-one from DDR, which ensures the sufficient overlapping of DMA transfer and PE array computation. The \emph{average execution time} of the sequence batch is estimated as the latency result. As shown in Table.~\ref{table:sota_e2e}, owing to the computation reduction of butterfly sparsity and the dataflow orchestration, our design achieves $23.69\times$ and $16.56\times$ latency reduction, as well as $6.37\times$ and $3.60\times$ energy efficiency over SpAtten\cite{SpAtten} and DOTA\cite{dota}. In terms of the SOTA butterfly work, our design achieves $1.17\times$ speedup and $3.36\times$ energy efficiency. It should be \textbf{emphasized} that our design is on the basis of a \textbf{\emph{reconfigurable dataflow architecture}} that has the flexibility from programmability to cover other NN operators or attention computation with other structured sparsity, unlike the dedicated ASIC architectures evaluated above.   

\section{Conclusion}
In this paper, we proposed multilayer dataflow orchestration on reconfigurable dataflow architecture supported with energy-efficient coarse-grained streaming parallelism, to accelerate butterfly sparsity computation for attention workloads. The experiments show that compared to NVIDIA Jetson Xavier NX with tensor cores, our design has a speedup of up to $14.34\times$ ($9.29\times$ on average) as well as $12.3\times$ energy efficiency advancement, owing to the computation efficiency of butterfly sparsity and the combined optimization of hardware architecture and dataflow mapping. In comparison with SOTA butterfly work, our design has $1.17\times$ speedup and $3.36\times$ energy efficiency improvement at the same peak performance.

\bibliographystyle{IEEEtran}
\bibliography{ref}

\end{document}